\documentclass[12pt,letterpaper]{article}

\usepackage{amsmath,amssymb,calc}
\usepackage{graphicx}
\usepackage{cite}

\newcommand\be{\begin{equation}}
\newcommand\ee{\end{equation}}
\newcommand{\bea}{\begin{eqnarray}}
\newcommand{\eea}{\end{eqnarray}}

\newcommand{\nn}{\nonumber}
\newcommand{\pd}{\partial}

\def\id{\protect{{1 \kern-.28em {\rm l}}}}

\def\id{\protect{{1 \kern-.28em {\rm l}}}}

\begin{document}

\begin{titlepage}
\begin{center}
\hfill \\
\vspace{1cm}
{\Large {\bf Dynamics of Inspiraling Dark Energy\\[3mm] }}

\vskip 1.5cm
{\bf Lilia Anguelova${}^a$\footnote{anguelova@inrne.bas.bg}, John Dumancic${}^b$\footnote{dumancjp@mail.uc.edu}, Richard Gass${}^b$\footnote{gassrg@ucmail.uc.edu}, L.C.R. Wijewardhana${}^b$\footnote{rohana.wijewardhana@gmail.com}\\
\vskip 0.5cm  {\it ${}^a$ Institute for Nuclear Research and Nuclear Energy,
Bulgarian Academy of Sciences, Sofia 1784, Bulgaria\\
${}^b$ Department of Physics, University of Cincinnati, Cincinnati, OH 45221, USA}}

\vskip 6mm

\end{center}

\vskip .1in
\vspace{0.5cm}

\begin{center} {\bf Abstract}\end{center}

\vspace{-1cm}

\begin{quotation}\noindent

We investigate the dynamics of a multifield dark energy model, which arises from certain rapid-turning and inspiraling trajectories in field space. We find the speed of sound $c_s$ of the dark energy perturbations around the background and show that $c_s$ is monotonically decreasing with time. Furthermore, it has a positive-definite lower bound that implies a certain clustering scale. We also extend the previously known background solution for dark energy to an exact solution that includes matter. This allows us to address the implications of our model for two cosmological tensions. More precisely, we argue that the $\sigma_8$ tension can be alleviated generically, while reducing the Hubble tension requires certain constraints on the parameter space of the model. Notably, a necessary condition for alleviating the Hubble tension is that the transition from matter domination to the dark energy epoch begins earlier than in $\Lambda$CDM.

\end{quotation}

\end{titlepage}

\eject

\tableofcontents

\section{Introduction}

Modern observations have shown that the current expansion of the Universe is accelerating. A simple explanation for that is provided by a cosmological constant. The corresponding standard model of cosmology, $\Lambda$CDM, gives a reasonably good phenomenological description of the evolution of the Universe. However, the more accurate the observational data becomes, the more discrepancies it leads to between our understanding of the early and of the late Universe. These inconsistencies are known as cosmological tensions. The most prominent of them are the so called Hubble tension and $\sigma_8$ tension.\footnote{The Hubble tension is a disagreement between the value of the Hubble constant, extracted from the CMB assuming $\Lambda$CDM \cite{A4B2etal}, and its value obtained from local observations in the late Universe \cite{Reissetal}. Similarly, the $\sigma_8$ tension is a disagreement between the determinations from the CMB (together with $\Lambda$CDM) \cite{A4B2etal}, on one hand, and from local observations \cite{Heymetal,Valetal,NV}, on the other, of the present-day value of the amplitude of linear matter fluctuations on a certain scale.} The existence of such tensions, as well as the unnaturally tiny value of the energy density that drives the present-day acceleration, motivate the search for alternative models of cosmology.
 
An appealing theoretical alternative to the cosmological constant is dynamical dark energy, arising from scalar field evolution. Models with multiple scalars are especially favored by recent arguments regarding the compatibility of effective field theories with quantum gravity \cite{OOSV,GK,OPSV,AP,BPR}. Such multifield models can lead to qualitatively different features, compared to the single scalar case, if they have background solutions with non-geodesic trajectories in field-space. Those trajectories are characterized by large turning rates. Cosmological models, relying on such a rapid turn regime, have been of interest for describing both inflation in the early Universe \cite{AB,SM,TB,BM,GSRPR,CRS,ACIPWW,APR,PSZ,FRPRW,LA,LA2_pbh,CR,IMS,ChG} and dark energy in the late Universe \cite{CDP,CDP2,ASSV,EASV,ADGW}. An important novel feature of such multifield dark energy is that, even for an equation-of-state parameter arbitrarily close to $-1$\,, it could be observationally distinguishable from a cosmological constant \cite{ASSV}. This raises great interest in studying models of that type. In \cite{ADGW} we derived a class of exact solutions, describing dark energy, in a certain two-field cosmological model. The background solutions of \cite{ADGW} are obtained by choosing the Poincar\'e disk as the field space and by using certain hidden symmetry. In addition, they are characterized by rapid-turning field-space trajectories, which are spiraling in toward the center of field space. Here we will investigate the dynamics of the resulting models.

Unlike a cosmological constant, the dark energy fields can fluctuate in spacetime. If the speed of sound of those perturbations is less than the speed of light, they can cluster on subhorizon scales and thus affect the large-scale structure of the Universe at late times. We study the behavior of perturbations of the dark energy scalars around the exact backgrounds of \cite{ADGW} and compute their speed of sound. We then show that the rapid-turn regime of our model is compatible with a reduced (compared to the speed of light) sound speed of these dark energy fluctuations. Furthermore, we prove that this speed of sound is a monotonically decreasing function of time, which has a positive-definite lower bound, everywhere in our parameter space. Interestingly, this bound is independent of any parameter, other than the turning rate, and becomes a pure constant for rapid turning. In addition, we show that the asymptotic regime of our solutions, in which the sound speed achieves its lower bound, is reached well within the first e-fold of accelerated expansion. This leads to a certain scale of dark energy clustering, which is a universal characteristic of the entire class of models.

We also extend the dark energy backgrounds of \cite{ADGW} to exact solutions that include matter. This allows us to consider the implications of our model for the $\sigma_8$ and Hubble tensions. Using our results for the dark energy equation-of-state parameter and perturbations' sound speed, we argue that the $\sigma_8$ tension should be alleviated generically, regardless of the choices of parameter values. We further show that a necessary condition for alleviating the Hubble tension, in our class of models, is that the transition from matter domination to the dark energy epoch begins earlier than in $\Lambda$CDM. This can be achieved in a large part of our parameter space, which opens the possibility to resolve both the $\sigma_8$ and Hubble tensions simultaneously.

It should be noted that the approximations, used in our considerations, are well-satisfied for small field values, i.e. field values below the Planck scale. This is a natural condition for the reliability of an effective field-theoretic description. However, it is often difficult to achieve in rapid turning models. Here we show that our solutions allow the combination of rapid turning with small field values, as well as with a reduced sound speed of the dark energy perturbations. Furthermore, the larger the turning rate is, the more well-satisfied the relevant approximations are. 

The organization of this paper is the following. In Section \ref{Prelim}, we review the basic ingredients of the dark energy solutions of \cite{ADGW}, as well as some of their characteristics that will be needed in the following. In Section \ref{PertAndSSpeed}, we consider dark energy perturbations around these backgrounds and derive a formula for the fluctuations' speed of sound. In Section \ref{BehaviorCs}, we investigate the dependence of this sound speed both on time and on parameter space. We prove that it is a monotonically decreasing function of time, with a positive-definite lower bound, everywhere in parameter space. Furthermore, we show that large turning rates occur in the same part of parameter space, which corresponds to a significantly reduced speed of sound. We also find the characteristic scale of dark energy clustering that results from the lower bound on the sound speed. In Section \ref{InclM}, we find a class of exact solutions of the background equations of motion, which generalizes the backgrounds of \cite{ADGW} in the presence of matter. We then argue that our model should alleviate the $\sigma_8$ tension everywhere in its parameter space. Furthermore, we show that, to alleviate the Hubble tension, one has to impose certain constraint on parameter space, which leads to earlier beginning of the dark energy era, compared to $\Lambda$CDM. Finally, in Section \ref{Discussion}, we summarize our results and discuss open issues that merit further investigation. The two Appendices contain technical details relevant for Section \ref{BehaviorCs}. In particular, in Appendix \ref{r_of_varphi} we show that the approximations, within which we have derived the speed of sound, are well-satisfied for small field values. And in Appendix \ref{w_DE} we prove that the equation-of-state parameter of our dark energy solutions reaches its asymptotic regime within the first e-fold of accelerated expansion.

\section{Preliminaries on dark energy background} \label{Prelim}
\setcounter{equation}{0}

We will study a class of dark energy models that arises from the minimal coupling of two scalar fields $\phi^I (x^{\mu})$ to Einstein gravity. This system is described by the following action:
\be \label{Action_gen}
S = \int d^4x \sqrt{-\det g} \left[ \frac{R}{2} - \frac{1}{2} G_{IJ} (\{\phi^I\}) \pd_{\mu} \phi^I \pd^{\mu} \phi^J - V (\{ \phi^I \}) \right] \,\,\, ,
\ee
where $g_{\mu \nu}$ with $\mu = 0,..,3$ is the metric on spacetime, $R$ is its scalar curvature and, finally, $G_{IJ}$ with $I = 1,2$ is the metric on the field-space with coordinates $\{\phi^I\}$\,. As usual, we will assume that the  cosmological background is given by a spacetime metric and scalar fields of the form:
\be \label{metric_g}
ds^2_g = -dt^2 + a^2(t) d\vec{x}^2 \qquad , \qquad \phi^I = \phi^I_0 (t) \quad ,
\ee 
where $a(t)$ is the scale factor. Then, the standard definition of the Hubble parameter is:
\be \label{Hp}
H (t) = \frac{\dot{a}}{a} \,\,\, , 
\ee
where $\dot{} \equiv d/dt$\,.

We will be interested in the dark energy solutions found in \cite{ADGW}. In that case, the field-space metric $G_{IJ}$ has the form:
\be \label{Gmetric}
ds^2_{G} = d\varphi^2 + f(\varphi) d\theta^2 \,\,\, ,
\ee
where for convenience we have denoted :
\be \label{Backgr_id}
\phi^1_0 (t) \equiv \varphi (t) \qquad , \qquad \phi^2_0 (t) \equiv \theta (t) \quad .
\ee
The background equations of motion that follow from the action (\ref{Action_gen}), when using (\ref{metric_g}) and (\ref{Gmetric}), are:
\be \label{ScalarEoMs}
\ddot{\varphi} - \frac{f'}{2} \dot{\theta}^2 + 3 H \dot{\varphi} + \pd_{\varphi} V = 0 \qquad , \qquad \ddot{\theta} + \frac{f'}{f} \dot{\varphi} \dot{\theta} + 3 H \dot{\theta} + \frac{1}{f} \pd_{\theta} V = 0  \quad ,
\ee
and
\be \label{EinstEq}
3H^2 = \frac{1}{2} \!\left( \dot{\varphi}^2 + f \dot{\theta}^2 \right) + V \quad ,
\ee
where ${}^{\prime} \equiv d/d\varphi$\,. To obtain genuine multifield models, that can lead to new effects compared to the single-field case, one needs to find solutions of (\ref{ScalarEoMs})-(\ref{EinstEq}), whose trajectories $(\varphi (t) , \theta(t))$ are not geodesics in the field space with metric (\ref{Gmetric}). Such field-space trajectories are characterized by a non-vanishing turning rate function.\footnote{We define this function formally in the next Section; see (\ref{Om_def}), as well as (\ref{Om}).} The rapid-turn regime, which corresponds to large deviations from geodesics, has attracted significant interest in recent years regarding novel models of cosmic acceleration for the purposes of both inflation \cite{AB,SM,TB,BM,GSRPR,CRS,ACIPWW,APR,PSZ,FRPRW,LA,LA2_pbh,CR,IMS,ChG} and late dark energy \cite{CDP,CDP2,ASSV,EASV,ADGW}. Note that a curved metric (\ref{Gmetric}) enables the existence of genuinely two-field (i.e., non-geodesic) background trajectories in field space even for potentials $V$ that do not depend on $\theta$\,, as is clear from (\ref{ScalarEoMs})-(\ref{EinstEq}); see also, for example, \cite{AB,SM,ABL}.

In \cite{ADGW} an exact dark-energy solution of (\ref{ScalarEoMs})-(\ref{EinstEq}) was found with the ansatz:
\be \label{th_d_om}
\dot{\theta} \equiv \omega = const \,\,\, .
\ee
To obtain this background solution, we also have to take the following expressions for the functions $f (\varphi)$ and $V (\varphi, \theta)$\,:
\be \label{fs}
f (\varphi) \, = \, \frac{8}{3} \,\sinh^2 \!\left( \sqrt{\frac{3}{8}} \,\varphi \right)
\ee
and
\be \label{Vs}
V (\varphi) \,\, = \,\, C_V \,\cosh^2 \!\left( \frac{\sqrt{6}}{4} \,\varphi \right) - \frac{4}{3} \,\omega^2 \quad , \quad C_V = const \,\,\, .
\ee
In addition, we have to require:
\be \label{restr}
3 C_V > 4 \omega^2 \,\,\, ,
\ee
in order to ensure the positive-definiteness of the scalar potential (\ref{Vs}). Note that, for $f(\varphi)$ as in (\ref{fs}), the field space with metric (\ref{Gmetric}) becomes the Poincar\'e disk, which is the simplest hyperbolic surface (for more details, see \cite{ABL}). The exact solution of \cite{ADGW}, resulting from the choices (\ref{th_d_om})-(\ref{restr}), is characterized by field-space trajectories $(\varphi (t), \theta(t))$ that are spirals toward the center of the Poincar\'e disk. Hence these trajectories deviate strongly from geodesics and, thus, lead to a genuine two-field (as opposed to an effective single-field) model. It should be underlined that this is a model of the late Universe only. Namely, it aims to describe only the present-day stage of accelerated expansion driven by dark energy. At earlier epochs other components (like matter, before that radiation etc.) dominate the evolution of the Universe and thus determine the behavior of the scale factor $a(t)$\,, whereas the dark energy scalars $\varphi$ and $\theta$ are negligible (and effectively frozen because of the large Hubble friction); see \cite{ASSV,EASV,ADGW}.\footnote{In Section \ref{InclM} we will show that it is actually easy to incorporate matter in the background solution of \cite{ADGW}. However, extending the model to include earlier epochs, like radiation domination and inflation, remains an interesting open problem.}  

Note that substituting (\ref{th_d_om}) and (\ref{Vs}) in (\ref{ScalarEoMs}) gives:
\be \label{Hf}
H = - \frac{f'}{3f} \dot{\varphi} \,\,\, .
\ee
From (\ref{EinstEq}), together with (\ref{th_d_om}) and (\ref{Hf}), one can obtain the following relation \cite{ADGW}:
\be \label{ph_d}
\dot{\varphi}^2 = \frac{V + \frac{1}{2} f \omega^2}{\frac{f'^2}{3f^2}-\frac{1}{2}} \,\,\, .
\ee
Substituting (\ref{fs}) and (\ref{Vs}) in (\ref{ph_d}) gives $\dot{\varphi}^2$ as a function of $\varphi$:
\be \label{ph_d_ph}
\dot{\varphi}^2 (\varphi) = \frac{2}{3} \,\sinh^2 \!\left( \frac{\sqrt{6}}{4} \,\varphi \right) \left[ \left( 3 C_V + 4 \omega^2 \right) \cosh^2 \!\left( \frac{\sqrt{6}}{4} \,\varphi \right) - 8 \omega^2 \right] \,\,\, .
\ee
This will be very useful below, since $\varphi (t)$ is a monotonically decreasing function in the solutions of \cite{ADGW}. So, instead of studying the $t$-dependence (and the limit $t \rightarrow \infty$) of the various relevant functions, we can study their $\varphi$-dependence (and the limit $\varphi \rightarrow 0$), which will simplify the analytical computations significantly.\footnote{One should keep in mind that during the earlier epochs, when the dark energy scalars are effectively frozen and $H$ is determined by the energy density of other components, relation (\ref{Hf}) and its consequences (\ref{ph_d})-(\ref{ph_d_ph}) do not apply. We will discuss this point further in Section \ref{InclM}.}

In view of the last paragraph, it will turn out that we do not need the explicit form of the solution of (\ref{ScalarEoMs})-(\ref{EinstEq}) found in \cite{ADGW}, in order to extract a number of general properties of the resulting model. Nevertheless, for completeness and more clarity, let us write down this exact background, for the scale factor $a(t)$ and the two scalars, namely:
\bea \label{Sol_om}
a (t) &=& \left( u^2 (t) - C_w^2 \right)^{1/3} \,\,\,\, , \nn \\
\varphi (t) &=& \sqrt{\frac{8}{3}} \,{\rm arccoth} \!\left( \sqrt{\frac{u^2 (t)}{C_w^2}} \,\,\right) \,\,\,\, , \nn \\
\theta (t) &=& \theta_0 + \omega t \,\,\,\, ,
\eea
where 
\be \label{Sol_u_om}
u (t) = C_1^u \sinh (\tilde{\kappa} t) + C_0^u \cosh (\tilde{\kappa} t) \qquad {\rm with} \qquad \tilde{\kappa} \equiv \frac{1}{2} \sqrt{3C_V - 4 \omega^2} \,\,\,\, ,
\ee
as well as $C_{0,1}^u,C_w,\theta_0=const$ and, in addition, the integration constants are related by the following constraint:
\be \label{Constr_cp}
\tilde{\kappa}^2 \!\left[ (C_0^u)^2 - (C_1^u)^2 \right] + 2 \omega^2 C_w^2 = 0 \,\,\,\, .
\ee
For a detailed discussion of the physical parameter space of (\ref{Sol_om})-(\ref{Constr_cp}), in which one is guaranteed to have $a(t),\dot{a}(t)>0$\,, we refer the reader to \cite{ADGW}.

\section{Perturbations and sound speed} \label{PertAndSSpeed}
\setcounter{equation}{0}

A distinguishing feature of dynamical dark energy, compared to a cosmological constant, is that it can have perturbations around the background solution. An important characteristic of the dynamics of the model is given by the speed of sound $c_s$ of these perturbations. If $c_s$ is smaller than the speed of light\footnote{We use standard theoretical units, in which the speed of light is $c=1$, as can be seen from (\ref{metric_g}).}, then the sound horizon of the perturbations is smaller than the particle horizon. As a consequence, for $c_s < 1$ there could be dark energy clustering on sub-horizon scales, which would affect structure formation. The goal of this Section is to derive the formula for the sound speed of the dark energy perturbations around the background solution of \cite{ADGW}. 

The perturbations around a given background $\phi_0^I (t)$ are defined via the following expansion of the scalars:
\be
\phi^I (t, \vec{x}) = \phi_0^I (t) + \delta \phi^I (t, \vec{x}) \,\,\, .
\ee
It is convenient to decompose the fluctuations $\delta \phi^I$ into components that are parallel and perpendicular to the field-space trajectory $(\phi^1_0 (t), \phi^2_0 (t))$ of the background solution.\footnote{Recall that, in the case of interest for us, the background trajectory $(\phi^1_0 (t), \phi^2_0 (t)) \equiv (\varphi (t) , \theta (t))$ is given explicitly by (\ref{Sol_om}), together with (\ref{Sol_u_om})-(\ref{Constr_cp}).} For that purpose, let us introduce the following orthonormal basis in field space:
\be \label{basisTN}
T^I = \frac{\dot{\phi}^I_0}{\dot{\sigma}} \quad , \quad N_I = (\det G)^{1/2} \epsilon_{IJ} T^J \quad , \quad \dot{\sigma}^2 = G_{IJ} \dot{\phi}^I_0 \dot{\phi}^J_0 \,\,\, .
\ee
Then the parallel and perpendicular components, respectively, are: 
\be
\delta \phi_{\parallel} = T_I \delta \phi^I \qquad , \qquad \delta \phi_{\perp} = N_I \delta \phi^I \quad .
\ee
Using (\ref{basisTN}), we can also define the turning rate of a field-space trajectory \cite{AAGP}:
\be \label{Om_def}
\Omega = - N_I D_t T^I \quad , \quad {\rm where} \quad D_t T^I \equiv \dot{\phi}_0^J \nabla_J T^I \,\,\, .
\ee
This quantity measures the deviation of the trajectory from a geodesic and thus plays an important role in rapid turning models, like those of \cite{ADGW,ASSV}.

To determine the speed of sound of the perturbations $\delta \phi^I$\,, we will follow the strategy used in \cite{ASSV} that captures the essential physics of the problem. In particular, we will neglect matter perturbations and gravitational backreaction, will use the rapid turn approximation and will consider only scales smaller than the sound horizon. Then, the leading terms in the equations of motion of the perturbations are (see \cite{ASSV} and references therein):\footnote{The $M_T^2$ term of (\ref{PertEoMs}) was neglected in the analogous eqs. (33)-(34) of \cite{ASSV}, because they assumed the slow roll approximation. However, the second slow roll parameter (defined as $\eta \equiv - \frac{\ddot{H}}{2 H \dot{H}}$) of the background solutions of \cite{ADGW} is not small, as was already observed in \cite{LA2}. Accordingly, we will see below that, indeed, $M_T^2$ cannot be neglected in our case. For further comparison with \cite{ASSV}, note that the primes in (33)-(34) there denote differentiation w.r.t. conformal time $\tau$\,. Transforming those equations to the form of our (\ref{PertEoMs}) involves not only the transformation $\tau \rightarrow t$\,, but also the use of the rapid turn approximation. Finally, in (\ref{PertEoMs}) and below, $\hat{k} = k/a$ is the physical wave number with $k$ being the comoving one.}
\bea \label{PertEoMs}
&&\delta \ddot{\phi}_{\parallel} + \left( \hat{k}^2 + M_T^2 \right) \delta \phi_{\parallel} \,= \,- 2 \Omega \delta \dot{\phi}_{\perp} \quad , \nn \\
&&\delta \ddot{\phi}_{\perp} + \left( \hat{k}^2 + M_{eff}^2 \right) \delta \phi_{\perp} \,= \,2 \Omega \delta \dot{\phi}_{\parallel} \quad ,
\eea
where
\be \label{MTMeff}
M_T^2 = T^I T^J V_{;IJ} \qquad , \qquad M_{eff}^2 = N^I N^J V_{;IJ} - \Omega^2 + \varepsilon H^2 {\cal R}
\ee
and
\be
V_{;IJ} \equiv \nabla_I \nabla_J V = \pd_I \pd_J V - \Gamma_{IJ}^K \pd_K V \,\,\, .
\ee
Here $\Gamma_{IJ}^K$ and ${\cal R}$ are, respectively, the Christoffel symbols and Ricci scalar of the metric $G_{IJ}$ and $\varepsilon \equiv - \frac{\dot{H}}{H^2}$ is the first slow roll parameter. Also, we have denoted the mass of the perturbation $\delta \phi_{\perp}$ by $M_{eff}^2$ for easier comparison with the literature.

Recall that for our background solutions the field-space metric is of the form (\ref{Gmetric}) and the scalar potential (\ref{Vs}) satisfies $\pd_{\theta} V = 0$\,. Hence the turning rate is given by \cite{LA}:
\be \label{Om}
\Omega = \frac{\sqrt{f}}{\left( \dot{\varphi}^2 + f \dot{\theta}^2 \right)} \,\dot{\theta} \,\pd_{\varphi} V \,\,\, ,
\ee
while the quantity $V_{NN} \equiv N^I N^J V_{;IJ}$ has the form \cite{LA}:
\be \label{VNN}
N^I N^J V_{;IJ} = \frac{ f \dot{\theta}^2 \pd_{\varphi}^2 V + \frac{f'}{2 f} \dot{\varphi}^2 \pd_{\varphi} V }{ ( \dot{\varphi}^2 + f \dot{\theta}^2 ) } \,\,\, .
\ee
Using (\ref{Gmetric}) and (\ref{basisTN}), we also compute:
\be \label{VTT}
T^I T^J V_{;IJ} = \frac{\dot{\varphi}^2 \pd_{\varphi}^2 V + \dot{\theta}^2 \frac{f'}{2} \pd_{\varphi}V }{( \dot{\varphi}^2 + f \dot{\theta}^2 ) } \,\,\, .
\ee
Substituting (\ref{Om})-(\ref{VTT}) in (\ref{MTMeff}) enables one to obtain explicit expressions for the masses of the perturbations in our case. We will investigate them in the next Section. Note that we will always neglect the $\varepsilon$ term in (\ref{MTMeff}) since the backgrounds of \cite{ADGW} are characterized by $\varepsilon <\!\!< 1$ and ${\cal R} = - \frac{3}{4}$\,. Notice also that substituting (\ref{th_d_om})-(\ref{Vs}) and (\ref{ph_d_ph}) inside (\ref{Om}) gives a (rather) non-trivial turning-rate function $\Omega (\varphi )$, as should be the case according to the discussion in the previous Section.\footnote{Recall that for (effectively) single-field models the turning rate vanishes identically.}

Now let us turn to finding the dispersion relation, which will enable us to extract the speed of sound. For that purpose, we consider solutions of the perturbations' equations of motion (\ref{PertEoMs}), whose dependence on time is of the form: \,$\delta \phi_{\perp} , \delta \phi_{\parallel} \sim e^{i \,\widetilde{\omega} \,t}$ \,.\footnote{The frequency $\widetilde{\omega}$ introduced here should not be confused with the constant $\omega$ in our ansatz (\ref{th_d_om}).} This leads to the following relation: 
\be \label{disp_rel}
\widetilde{\omega}^4 - \widetilde{\omega}^2 \!\left( 2 \hat{k}^2 + M_T^2 + M_{eff}^2 + 4 \Omega^2 \right) + \left( \hat{k}^2 + M_T^2 \right) \!\left( \hat{k}^2 + M_{eff}^2 \right) = 0 \,\,\,\, .
\ee
The solutions of (\ref{disp_rel}) are:
\be \label{om_t_pm}
\widetilde{\omega}_{\pm}^2 = \hat{k}^2 + \frac{M_T^2 + M_{eff}^2 + 4 \Omega^2}{2} \pm \sqrt{\left( \frac{M_T^2 + M^2_{eff} + 4 \Omega^2 }{2} \right)^{\!\!2} + 4 \Omega^2 \hat{k}^2 - M_T^2 M_{eff}^2} \,\,\,\, .
\ee
Considering scales, such that $\hat{k}^2 <\!\!< M_T^2 + M_{eff}^2 + 4 \Omega^2$\,, and expanding (\ref{om_t_pm}) to leading order gives:
\bea \label{om_pm}
\widetilde{\omega}_{\pm}^2 \!&=& \!\hat{k}^2 + \frac{M_T^2 + M_{eff}^2 + 4 \Omega^2}{2} \pm \left[ \left( \frac{M_T^2 + M^2_{eff} + 4 \Omega^2 }{2} \right)^{\!\!2} - M_T^2 M_{eff}^2 \right]^{\!\frac{1}{2}} \nn \\
\!&\pm \!& \frac{1}{2} \,\frac{4 \Omega^2 \hat{k}^2}{ \left[ \left( \frac{M_T^2 + M^2_{eff} + 4 \Omega^2 }{2} \right)^{\!\!2} - M_T^2 M_{eff}^2 \right]^{\frac{1}{2}} } \, + \, {\cal O} (\hat{k}^4) \quad .
\eea
This modifies the result of \cite{ASSV}, for scales larger than the Compton wavelength of the $\widetilde{\omega}_+$ mode, in the presence of non-negligible $M_T^2$\,. Note that from (\ref{om_pm}) one can obtain a light mode with a dispersion relation of the form:
\be \label{om_cs_k}
\widetilde{\omega}^2_- \approx c_s^2 \hat{k}^2 + {\cal O} (\hat{k}^4) \,\,\, ,
\ee
where
\be \label{cs_2}
c_s^{-2} \approx 1 + \frac{4 \Omega^2}{M_T^2 + M_{eff}^2} \,\,\, ,
\ee
by requiring that $M_{eff}^2 <\!\!< 1$ or even just:
\be \label{Ineq}
M_T^2 M_{eff}^2 \,<\!\!< \,\frac{1}{4} \left( M_T^2 + M_{eff}^2 + 4 \Omega^2 \right)^2 \,\,\, .
\ee
We will see below that, in our parameter space, rapid turning naturally corresponds to small $M_{eff}^2$\,. However, satisfying (\ref{Ineq}) is enough to ensure that the speed of sound of the $\widetilde{\omega}_-$ mode in (\ref{om_pm}) is well-approximated by the formula (\ref{cs_2}). Again, by taking $M_T^2=0$ in (\ref{cs_2}), we recover the result of \cite{ASSV}.

\section{Behavior of sound speed} \label{BehaviorCs}
\setcounter{equation}{0}

In this Section we will investigate the behavior of the speed of sound, determined by (\ref{cs_2}), for dark energy perturbations around the background solution of \cite{ADGW}. A straightforward way of addressing this problem would be to substitute the explicit form of $\varphi(t)$ given in (\ref{Sol_om})-(\ref{Constr_cp}), together with (\ref{th_d_om})-(\ref{Vs}), inside (\ref{Om})-(\ref{VTT}), in order to obtain explicitly the function $c_s (t)$ in our case. However, the resulting expression is quite messy and unmanageable. A more elegant approach is to use the fact that for our dark energy solutions $\varphi (t)$ is monotonically decreasing \cite{ADGW}. So we can use (\ref{ph_d_ph}) to obtain and investigate $c_s (\varphi)$\,, while remembering that increasing $t$ corresponds to decreasing $\varphi$\,. An additional benefit of this approach is that the results are manifestly independent on the values of the integration constants of the solution (\ref{Sol_om})-(\ref{Constr_cp}), except on those that also enter the potential (\ref{Vs}). Hence, only the latter two constants can lead to parameter-space dependence of the conclusions, reached by studying the function $c_s (\varphi)$\,. 

To gain insight into the behavior of the sound speed, we will begin by considering the large-$t$ limit or, equivalently, the limit of small $\varphi$\,. Later on, we will show that this limit gives a good description of the physics of the model for any $\varphi$ below the Planck scale. We will study both the time-dependence of $c_s$ and its dependence on parameter space. 

Recall that dark energy perturbations can affect structure formation only if their speed of sound is significantly suppressed compared to $c_s=1$\,. So let us first address the question whether a suppressed $c_s (\varphi)$ is compatible with the rapid-turning regime, in the parameter space of our class of models. For that purpose, we need to compare the dimensionless turning rate $\Omega/H$ and the sound speed $c_s$\,. Substituting (\ref{th_d_om})-(\ref{Vs}) and (\ref{ph_d_ph}) inside (\ref{Hf}) and (\ref{Om})-(\ref{VTT}), and then using (\ref{MTMeff}) and (\ref{cs_2}), we find for $\varphi \rightarrow 0$:
\be \label{Om_H_lead}
\left( \frac{\Omega}{H} \right)^{\!2} \quad \rightarrow \quad \frac{9 \omega^2}{3C_V - 4 \omega^2} + {\cal O} (\varphi^2)
\ee
and
\be \label{cs_lead}
c_s^{-2} \quad \rightarrow \quad 1 + \frac{8 \omega^2}{3C_V - 2\omega^2} + {\cal O} (\varphi^2) \quad ,
\ee
where we have used:
\be \label{Msq_as}
M_T^2 \,\, \rightarrow \,\, \frac{3}{4} C_V + {\cal O} (\varphi^2) \quad , \quad M_{eff}^2 = V_{NN} - \Omega^2 \,\, \rightarrow \,\, \frac{3}{4} C_V - \omega^2 + {\cal O} (\varphi^2) \quad .
\ee
Note that, due to (\ref{restr}), the expression $\frac{8 \omega^2}{3C_V - 2\omega^2}$ in the leading term of
the expansion 
(\ref{cs_lead}) is positive-definite, as should be the case in order to guarantee that $c_s^{-2} \ge 1$ (and thus $c_s^2 \le 1$) in accordance with (\ref{cs_2}). The nontrivial question is whether $c_s^{-2}$ can be appreciably larger than 1 in our class of dark energy models. Comparing (\ref{Om_H_lead}) and (\ref{cs_lead}), one can see that, indeed, large turning rates are perfectly compatible with large $c_s^{-2}$\,. 

Let us illustrate this point on a couple of numerical examples. Taking $\omega = 4$ and $C_V = 22$\,, as in the plots inside \cite{ADGW}, we obtain from (\ref{Om_H_lead}) and (\ref{cs_lead}) that $(\Omega / H)^2 \approx 72$ and $c_s^{-2} \approx 4.76$\,, implying that $c_s^2 \approx 0.21$ for small $\varphi$\,. In other words, we have a large turning rate together with a significantly suppressed (compared to $c_s^2 = 1$) speed of sound. Note that in this example we also have $\Omega^2 \approx 16$\,, $M_T^2 \approx 16.5$ and $M_{eff}^2 \approx 0.5$\,, for small $\varphi$\,. Thus, the inequality (\ref{Ineq}) is satisfied to a very good degree of accuracy:
\be
M_T^2 M_{eff}^2 \approx 8.25 \quad <\!\!< \quad \frac{1}{4} \left( M_T^2 + M_{eff}^2 + 4 \Omega^2 \right)^2 \approx 1640.25 \quad .
\ee
Although in this particular case $M_{eff}^2$ is not very small, it is clear from comparing (\ref{Om_H_lead}) and (\ref{Msq_as}) that larger turning rates correspond to smaller $M_{eff}^2$\,. In other words, the rapid-turn regime is naturally compatible with $M_{eff}^2 <\!\!< 1$\,. Note also that it is not necessary to have $\omega > 1$\,, as long as $\omega \neq 0$\,. Indeed, taking for instance $\omega = 0.1$ and $C_V = 1.35 \times 10^{-2}$\,, we obtain $(\Omega / H)^2 \approx 180$ and $c_s^2 \approx 0.20$\,. Furthermore, in this second example we have $\Omega \approx 10^{-2}$\,, $M_T^2 \approx 1.01 \times 10^{-2}$ and $M_{eff}^2 \approx 1.25 \times 10^{-4}$\,, implying again that the inequality (\ref{Ineq}) is satisfied by two orders of magnitude but now together with $M_{eff}^2 <\!\!< 1$\,.

We will investigate more systematically the region of parameter space, that gives both large turning rates and suppressed sound speeds, in Subsection \ref{Dep_on_par_space}. However, now that we have seen that these two conditions are compatible with each other, it will be useful to study first the dependence of $c_s$ on time. In the next Subsections we will show that the function $c_s (t)$\,, determined by (\ref{cs_2}), is monotonically decreasing {\it everywhere} in our parameter space and for {\it any} \,$\varphi \ge 0$\,. And, furthermore, it has a positive-definite lower bound. These results will be important in the later considerations.

\subsection{Time dependence}

We will study the time dependence of $c_s^2 (t)$ by investigating $c_s^2 (\varphi)$\,, as explained above. To obtain the sound speed as a function of $\varphi$\,, let us substitute (\ref{th_d_om}) and (\ref{ph_d_ph}) in (\ref{Om})-(\ref{VTT}). Using these results, together with (\ref{fs})-(\ref{Vs}), inside (\ref{MTMeff}) and then in (\ref{cs_2}), we obtain:
\be \label{cs_f}
c_s^{-2} (\varphi) = 1 + \frac{48 \omega^2 C_V \cosh^2 \!\hat{\varphi}}{Q (\hat{\varphi})} \quad ,
\ee
where for convenience
\be \label{phi_hat}
\hat{\varphi} \equiv \frac{\sqrt{6}}{4} \,\varphi
\ee
and
\bea
Q (\hat{\varphi}) &=& \left( 27 C_V^2 + 72 C_V \omega^2 + 48 \omega^4 \right) \cosh^6 \!\hat{\varphi} - \left( 112 \omega^4 + 96 C_V \omega^2 + 9 C_V^2 \right) \cosh^4 \!\hat{\varphi} \nn\\
&+& \left( 12 C_V \omega^2 + 80 \omega^4 \right) \cosh^2 \!\hat{\varphi} - 16 \omega^4 \,\,\, .
\eea

Note that for large $\varphi$ (i.e., at early times) the expression $\frac{48 \omega^2 C_V \cosh^2 \!\hat{\varphi}}{Q (\hat{\varphi})}$ in (\ref{cs_f}) behaves as $\cosh^{-4} \!\hat{\varphi}$ and thus $c_s^2 \rightarrow 1$\,. On the other hand, for small $\varphi$ (i.e., at late times), we have the expansion:
\be \label{cs_subl}
c_s^{-2} (\varphi) \,\, = \,\, 1 \,+ \,\frac{8 \omega^2}{3C_V - 2\omega^2} \,- \,\frac{3}{2} \,\frac{ \omega^2 \left( 15 C_V + 16 \omega^2 \right) }{\left( 3 C_V - 2 \omega^2 \right)^2} \,\varphi^2 \,+ \,{\cal O} (\varphi^4) \,\,\, ,
\ee
whose leading term is given in (\ref{cs_lead}). In view of (\ref{restr}), the coefficient of the $\varphi^2$ term in (\ref{cs_subl}) is negative-definite. Hence with time (i.e., with decreasing $\varphi$) $c_s^{-2}$ increases, tending to the constant leading term. In other words, $c_s^2$ decreases with time, tending to a lower bound. It should be noted that this lower bound is positive-definite, again due to (\ref{restr}). Hence, in our case there is no gradient instability anywhere in the physical parameter space, unlike in the model of \cite{ASSV,EASV}.

The consideration of the small $\varphi$ limit gave us a useful insight in the behavior of the speed of sound. With this in mind, we will show in the following that $c_s^2 (t)$ is a monotonically decreasing function for any $t\ge t_0$\,, where $t_0$ is the initial moment of the spatial expansion described by our model.\footnote{In this Section, $t_0 > 0$ is the time, when the dark-energy-driven accelerated expansion begins. In the next Section, we will extend the model to include matter, thus allowing us to have an initial moment $t_0$ during the earlier epoch of matter domination.} To achieve this, let us differentiate (\ref{cs_f}) with respect to $\varphi$\,. The result is:
\be \label{cs_sq_der}
\left( c_s^{-2} (\varphi) \right)' = \frac{A (\varphi)}{B (\varphi)} \quad ,
\ee
where
\be \label{AP}
A (\varphi) = -24 \sqrt{6} \,C_V \,\omega^2 \cosh \!\hat{\varphi} \,\sinh \!\hat{\varphi} \,P(\varphi)
\ee
with
\be
P(\varphi) = (144 C_V \omega^2 + 54 C_V^2 + 96 \omega^4) \cosh^6 \!\hat{\varphi} - (112 \omega^4 + 96 C_V \omega^2 + 9 C_V^2) \cosh^4 \!\hat{\varphi} + 16 \omega^4 \,\, ,
\ee
while $B (\varphi)$ is a messy expression that is a perfect square.\footnote{The expression for $B$ is not particularly illuminating and so we will not write it down explicitly.} Hence the sign of $\left( c_s^{-2} (\varphi) \right)'$ is determined by the sign of $A (\varphi)$\,. Furthermore, in our case $\varphi \ge 0$ always, since it is the radial coordinate on the Poincar\'e disk \cite{ADGW}. Therefore, the expression in (\ref{AP}) has a definite sign, when $P (\varphi)$ does.

Now we will show that $P (\varphi) \ge 0$\,, implying that $A (\varphi) \le 0$\,, for any $\varphi \ge 0$\,. To achieve this, it is useful to rewrite $P$ as: 
\be \label{Multiplier}
P (\varphi) = (48 C_V \omega^2 + 45 C_V^2) \cosh^4 \!\hat{\varphi} + (144 C_V \omega^2 + 54 C_V^2) \cosh^4 \!\hat{\varphi} \,\sinh^2 \!\hat{\varphi} + 16 \omega^4 T (\varphi) \,\,\, ,
\ee
where
\be \label{T_expr}
T (\varphi) = 6 \cosh^4 \!\hat{\varphi} \,\sinh^2 \!\hat{\varphi} - \cosh^4 \!\hat{\varphi} + 1 \,\,\, .
\ee
Clearly, the first two terms in (\ref{Multiplier}) are non-negative. To show that the last term is also non-negative, let us rewrite the expression in (\ref{T_expr}) as:
\be \label{T_expr_p}
T (\varphi) = 6 \cosh^6 \!\hat{\varphi} - 7 \cosh^4 \!\hat{\varphi} + 1 \,\,\, .
\ee
Now it is easy to see that $T(\varphi) \ge 0$\,. Indeed, differentiating (\ref{T_expr_p}) with respect to $\varphi$\,, we obtain:
\be
T' (\varphi) = \sqrt{6} \cosh^3 \!\hat{\varphi} \,( 2\sinh \!\hat{\varphi} + \sinh^3 \!\hat{\varphi}) \,\,\, .
\ee
Hence $T' \ge 0$ always due to $\hat{\varphi} \ge 0$\,. Thus $T (\varphi)$ is monotonically increasing. In addition, at $\varphi = 0$ we have $T = 0$\,. Therefore, $T \ge 0$ for any $\varphi \ge 0$\,. Thus every term in (\ref{Multiplier}) is non-negative. This implies, in view of (\ref{AP}), that $ A (\varphi) \le 0$ for any $\varphi \ge 0$\,. Hence from (\ref{cs_sq_der}) we have:
\be
\left( c_s^{-2} (\varphi) \right)' \le 0 \quad \,{\rm for} \,\quad \forall \varphi \ge 0 \quad .
\ee
Therefore, $c_s^{-2} (\varphi)$ is monotonically decreasing, which implies that $c_s^{-2} (t)$ is monotonically increasing. Equivalently, $c_s^2 (t)$ is a monotonically decreasing function.

Note that this result for the behavior of (\ref{cs_2}) is valid everywhere in our parameter space since its derivation did not use any other condition than $\varphi \ge 0$\,, which is a defining feature for the choice of field-space metric that leads to the solutions of \cite{ADGW}. However, the formula (\ref{cs_2}) itself was derived under an approximation. We will explore the implications of the latter in the next Subsections.

\subsection{Clustering scale} \label{Clust_scale}

We saw above that $c_s^2 (t)$ is a monotonically decreasing function. Hence the limit $t \rightarrow \infty$ (or, equivalently, the limit $\varphi \rightarrow 0$) in (\ref{cs_lead}) gives its lower bound. Let us now investigate the dependence of that bound on the turning rate. We will see that rapid turning implies a reduction of the sound speed and that, furthermore, it leads to a characteristic clustering scale.

For convenience, let us begin by recalling the leading behaviors, for small $\varphi$\,, of the relevant functions from (\ref{Om_H_lead}) and (\ref{cs_lead}), namely:
\be \label{large_eta}
\eta_{\perp}^2 \equiv \left( \frac{\Omega}{H} \right)^2 \approx \frac{9 \omega^2}{ 3 C_V - 4 \omega^2}
\ee
and
\be \label{large_cs}
c_s^{-2} \approx 1 + \frac{8 \omega^2}{ 3 C_V - 2 \omega^2} \,\,\,\, ,
\ee
respectively. Now, from (\ref{large_eta}), we have:
\be
C_V \approx \frac{\omega^2}{3} \left( 4 + \frac{9}{\eta_{\perp}^2} \right) \,\,\, .
\ee
Substituting this in (\ref{large_cs}) gives:
\be
c_s^{-2} \approx 1 + \frac{8}{\left( 2+\frac{9}{\eta_{\perp}^2} \right)} \,\,\,\, .
\ee
Hence, for $\eta_{\perp}^2 >\!\!> 1$ (i.e. rapid turning), we obtain: 
\be
c_s^{-2} \approx 5 \,\,\,\, .
\ee
In other words, for large turning rates, the lower bound on the speed of sound is:
\be \label{cs_b}
c_s^2 = 0.2 \,\,\,\, .
\ee
Note that this result does not depend at all on the choices of parameter values in our background solutions. Hence the bound (\ref{cs_b}) is {\it universal} \,for our class of multifield dark energy models.

Although the above conclusion was reached in the limit of small $\varphi$\,, its implications are not restricted to the $t \rightarrow \infty$ limit for the following reason. Recall that an effective field-theoretic description is most reliable below the Planck scale. In our context, this means considering small field values, i.e. the field-space region with $\varphi <\!\!< 1$ or at least $\varphi < 1$ \,.\footnote{We are working in units, in which the Planck mass is $M_p = 1$\,, as can be seen from (\ref{Action_gen}). In this context, large field values means $\varphi >\!\!> 1$ or at least $\varphi > 1$\,.} Since our solution for $\varphi (t)$ is monotonically decreasing, choosing an initial value $\varphi_0 \equiv \varphi|_{t=t_0}$\,, such that $\varphi_0 < 1$\,, ensures that the entire trajectory of the solution will lie in the region of small field values. In fact, the latter region was tacitly assumed in \cite{ADGW}; thus the numerical choices of the constants, used for the plots in the Figure there. So assuming an initial condition in the small field region, in order to ensure the reliability of our effective model\footnote{We will see later on that small field values are also required to satisfy the inequality (\ref{Ineq}) in the case of our background solutions.}, we can conclude that (\ref{cs_b}) approximates well the value of the sound speed throughout the entire evolution, described by our solution. 

In view of the above, using $c_s \approx 0.447$\,, we find that the sound horizon for dark energy perturbations, in our class of models, is of order:
\be \label{rs}
r_s \,\approx \,c_s \int_{0}^{t_*} \frac{dt}{a} \,\approx \,6.37 \,{\rm Gpc} \,\,\, ,
\ee
where \,$\int_{0}^{t_*} dt / a \approx 14.26 \,\,{\rm Gpc}$ \,is the comoving radius of the observable Universe with $t=0$ being the moment of the Big Bang and $t=t_*$ - the present moment. This determines a characteristic spatial scale such that, on scales of that order or larger, structure formation is enhanced by dark energy clustering. In principle, the effects of such clustering are quite difficult to detect. Nevertheless, they could be discerned for instance by analyzing cross-correlations between observational data on the integrated Sachs-Wolfe effect in the CMB, on one hand, and galaxy surveys, on the other; see for example \cite{HS}. Although the precision of data regarding the largest spatial scales is limited by cosmic variance, clustering on scales of order (\ref{rs}) might just be accessible to future observations.

\subsection{Dependence on parameter space} \label{Dep_on_par_space}

Our considerations so far have not depended on the choice of particular parameter values, although they did rely on certain approximations. Now we will address the question in what part of parameter space these approximations are well satisfied. Specifically, we will investigate in more detail the overlap between the region of rapid turning and that of reduced sound speed, as well as will show that (\ref{cs_2}) is a good approximation in that same part of parameter space.

\begin{figure}[t]
\begin{center}
\hspace*{-0.2cm}
\includegraphics[scale=0.35]{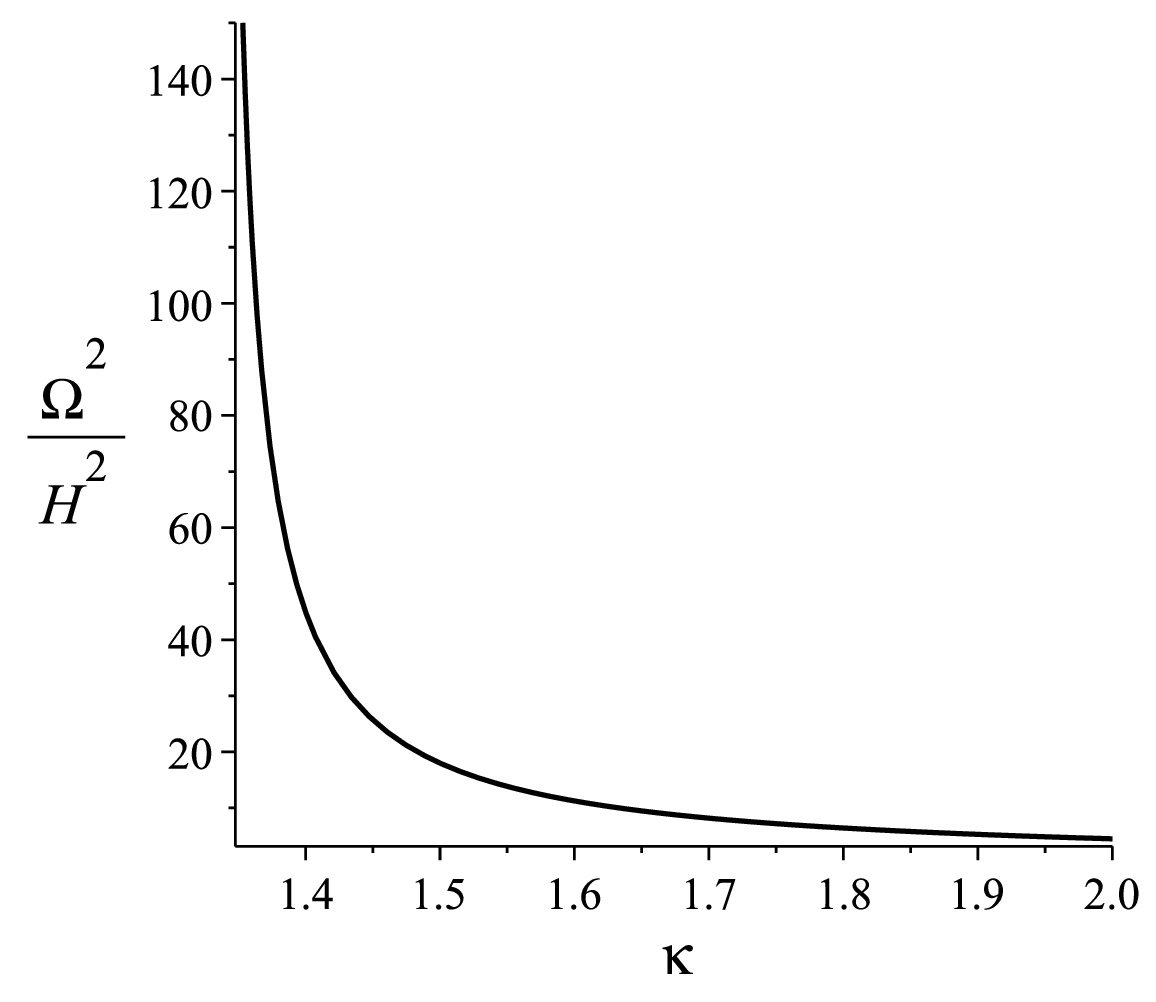}
\hspace*{0.4cm}
\includegraphics[scale=0.343]{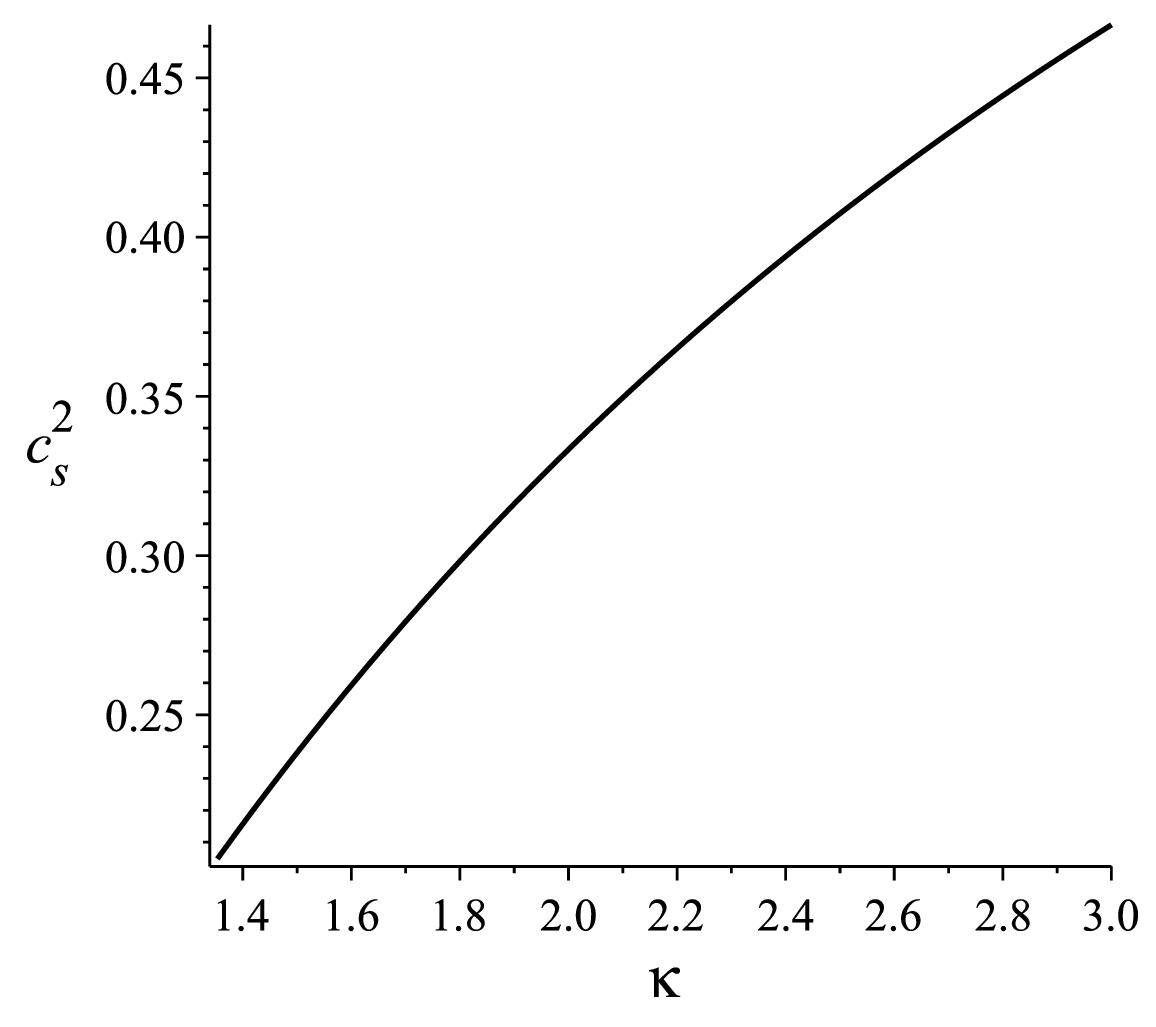}
\end{center}
\vspace{-0.7cm}
\caption{{\small Plots of the dimensionless turning rate $\eta_{\perp}^2 = \left( \frac{\Omega}{H} \right)^2$ and the speed of sound $c_s^2$ as functions of the parameter $\kappa$\,, according to (\ref{eta_cs_k}).}}
\label{ParSpace}
\vspace{0.1cm}
\end{figure}

With the previous Subsection in mind, we begin by focusing on small field values. At the end of this Section, we will  discuss large field values as well. For convenience, let us introduce the notation:
\be \label{k_def}
\kappa \equiv \frac{C_V}{\omega^2} \,\,\,\, .
\ee
Using this in (\ref{large_eta})-(\ref{large_cs}), one has for small $\varphi$:
\bea \label{eta_cs_k}
\eta_{\perp}^2 &=& \frac{9}{3 \kappa - 4} \,\,\,\, , \nn \\
c_s^{-2} &=& 1 + \frac{8}{3 \kappa - 2} \,\,\,\, . 
\eea
We plot the functions $\eta_{\perp}^2 (\kappa)$ and $c_s^2(\kappa)$\,, determined by (\ref{eta_cs_k}), in Figure \ref{ParSpace}. 
Clearly, smaller values of $\kappa$ give simultaneously larger $\eta_{\perp}^2$ and smaller $c_s^2$\,. In particular, the same region of parameter space, that corresponds to turning rates $\eta_{\perp}^2 \gtrsim 20$\,, also gives significantly reduced sound speeds $c_s^2 \lesssim 0.25$\,. Note that (\ref{restr}) implies $\kappa > \frac{4}{3}$\,. It is worth emphasizing that the above considerations depend only on the parameters in the potential (\ref{Vs}), but not on any of the rest of the integration constants in the explicit solutions of \cite{ADGW}. 

\begin{figure}[t]
\begin{center}
\hspace*{-0.4cm}
\includegraphics[scale=0.34]{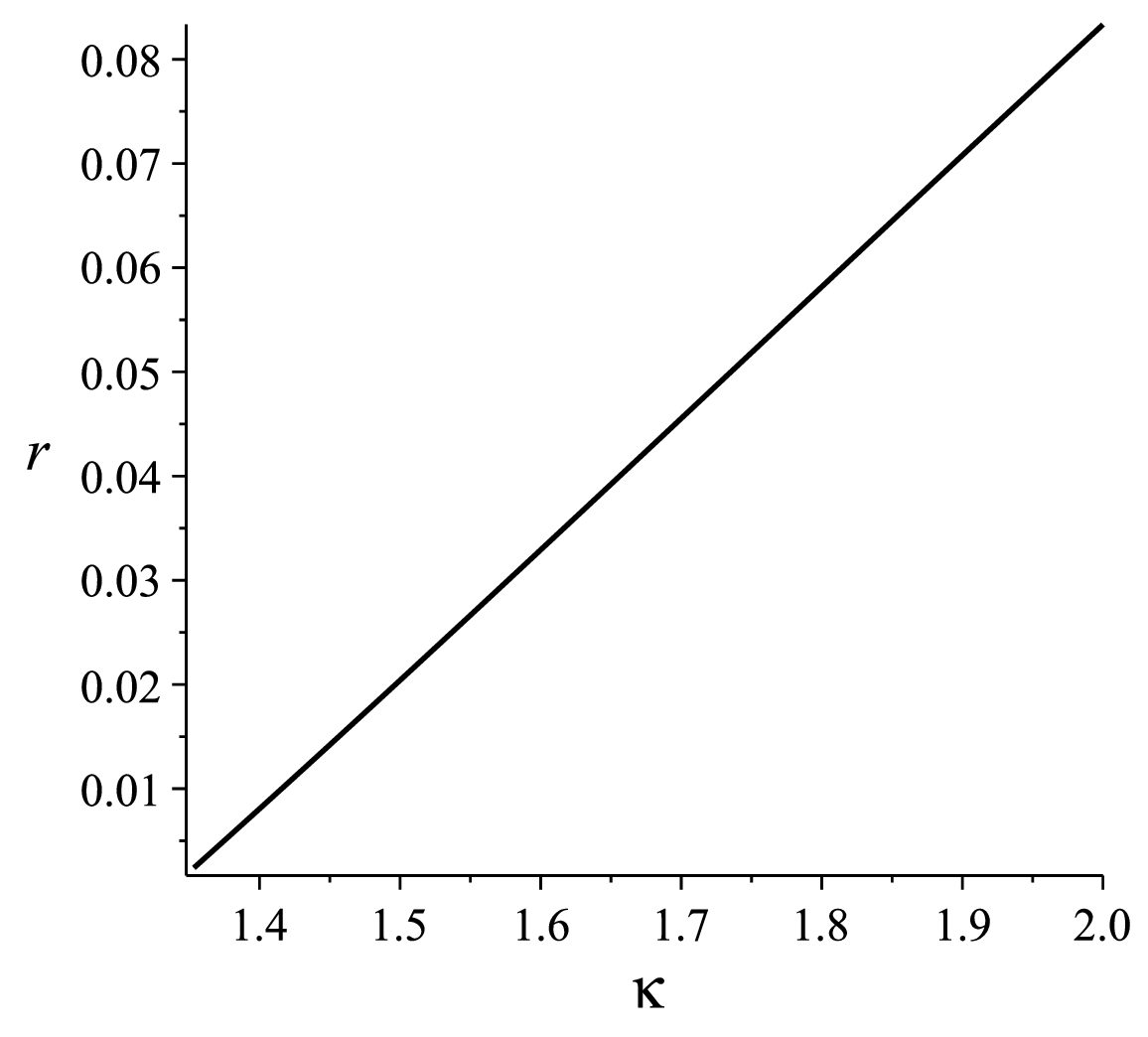}
\hspace*{0.6cm}
\includegraphics[scale=0.34]{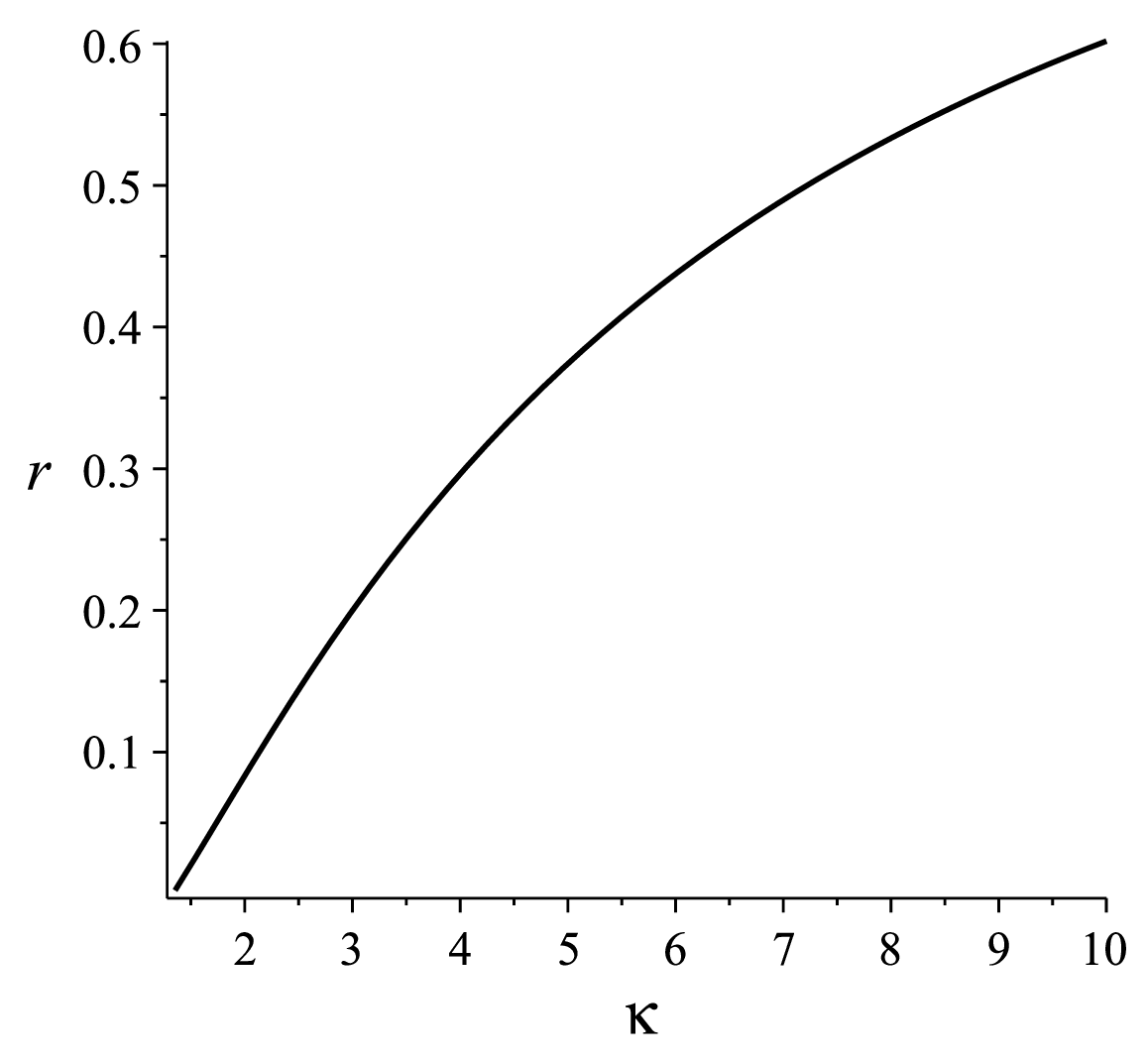}
\end{center}
\vspace{-0.7cm}
\caption{{\small Plots of the function $r (\kappa)$ given in (\ref{r_as}).}}
\label{r-k}
\vspace{0.1cm}
\end{figure}

Now we turn to verifying that the same region of parameter space, which gives large turning rates and suppressed sound speeds, is compatible with the approximations that lead to the formula (\ref{cs_2}). To investigate the inequality (\ref{Ineq}), let us consider the ratio:
\be \label{r_def}
r \equiv \frac{4 M_T^2 M_{eff}^2}{\left( M_T^2 + M_{eff}^2 + 4 \Omega^2 \right)^2} \,\,\,\, .
\ee
Using the leading order expressions (\ref{Msq_as}), together with $\Omega^2 \approx \omega^2 + {\cal O} (\varphi^2)$ \cite{ADGW} and the definition (\ref{k_def}), we find that for small $\varphi$:
\be \label{r_as}
r \approx \frac{\kappa (3 \kappa - 4)}{3 \left( \kappa + 2 \right)^2} \,\,\,\, .
\ee
Clearly, this expression implies that $r \rightarrow 0$ for $\kappa \rightarrow \frac{4}{3}$\,, whereas $r \rightarrow 1$ for $\kappa \rightarrow \infty$\,. It is also easy to realize that $r < 1$ for any $\kappa \in (\frac{4}{3} , \infty)$\,. We illustrate the dependence of the ratio $r$ on the parameter $\kappa$ in Figure \ref{r-k}. On the left of the Figure, we have chosen a smaller range for $\kappa$, in order to make explicit that $r <\!\!< 1$ in the part of parameter space, relevant for large turning rates and reduced $c_s^2$\,, while on the right of the Figure we have taken a larger range for $\kappa$ to exhibit better the overall shape of the function. By comparing Figures \ref{ParSpace} and \ref{r-k}, one can see that the inequality $r <\!\!< 1$\,, and thus (\ref{Ineq}), is well-satisfied numerically in the entire range, for which $\eta_{\perp}^2 \gtrsim 10$\,. Finally, note that for small $\varphi$ we have $M_{eff}^2 \approx \omega^2 (\frac{3}{4} \kappa - 1)$\,, which implies that $M_{eff}^2 \rightarrow 0$ in the limit $\kappa \rightarrow \frac{4}{3}$ that corresponds to large turning rates. So, to summarize, the use of the approximate formula (\ref{cs_2}) is well-justified in our case. 

In the above considerations, we assumed that the trajectories of our solutions lie in the small field region, as is preferable for the reliability of the effective field-theoretic description. However, depending on the initial condition $\varphi_0$\,, our solutions could have large field values at early times. Also, there are a number of large-field models in the literature, especially in the context of rapid turning. So it is worth investigating the behavior of the ratio (\ref{r_def}), as well as of $M_{eff}^2$\,, for any $\varphi$\,. In Appendix \ref{r_of_varphi}, we show that $r$ is small for $\varphi < 1$\,, while $r \sim {\cal O} (1)$ for $\varphi > 1$\,. Thus the approximation (\ref{Ineq}) breaks down for large field values. Furthermore, $M_{eff}^2$ increases with $\varphi$\,, such that the ratio $M_{eff}^2/\omega^2$ becomes large for $\varphi>1$\,. This affirms the choice of initial condition $\varphi_0 < 1$\,, or even $\varphi_0 <\!\!< 1$\,, in order to ensure the validity of the considerations of this Section throughout the entire evolution of a given solution. Note, however, that all solutions of \cite{ADGW} reach very fast, with increasing number of e-folds, their asymptotic regime that is characterized by small $\varphi$\,, as we show in Appendix \ref{w_DE}.\footnote{Interestingly, although in this regime our equation-of-state parameter is very close to $-1$ as in \cite{ASSV}, our reduced $c_s$ has different characteristics than theirs; we will discuss those differences in Section \ref{Discussion}.}

\section{Inclusion of matter} \label{InclM}
\setcounter{equation}{0}

So far we have considered backgrounds, which describe only dark energy. Now we will show that matter can be easily incorporated into the exact solutions of \cite{ADGW}. This will allow us to address the implications of our model for certain cosmological tensions, which arise from assuming the $\Lambda$CDM model of cosmology. It should be stressed that resolving completely these tensions is a well known theoretical challenge, which is still an open problem despite an enormous amount of literature on the topic. Our goal here will not be to present a complete solution\footnote{That would take us too far afield from the main subject of this paper - the speed of sound of the dark energy perturbations - and would certainly require an involved separate investigation.}, but merely to argue that our class of models holds promise for alleviating both the Hubble and $\sigma_8$ tensions. This should provide strong incentive for further studies of multifield dark energy.

To set the stage, let us note that, in the presence of matter, the equations of motion of the scalars remain the same as (\ref{ScalarEoMs}), while the Friedman equation (\ref{EinstEq}) is modified to:
\be \label{EinstEq_m}
3H^2 = \frac{1}{2} \!\left( \dot{\varphi}^2 + f \dot{\theta}^2 \right) + V + \rho_m \quad ,
\ee
where $\rho_m$ is the matter density, given as usual by:
\be \label{rho_m}
\rho_m = \frac{\rho_{m_0}}{a^3} \qquad , \qquad \rho_{m_0} = const \quad .
\ee
Since we are aiming to generalize the dark energy solutions of \cite{ADGW}, we will take again the same $f$ and $V$ as in (\ref{fs}) and (\ref{Vs}), respectively. 

The above equations of motion can be solved analytically in a manner very similar to the considerations of \cite{ADGW}. The basic idea is to use techniques from the Noether symmetry method and, more specifically, a hidden symmetry found in \cite{ABL}, in order to simplify greatly the system given by (\ref{ScalarEoMs}) and (\ref{EinstEq_m}), when $f$ and $V$ are as in (\ref{fs}) and (\ref{Vs}) respectively. For more details on the Noether symmetry method, see \cite{CdR,CMRS,CNP,CDeF} (albeit in the context of extended theories of gravity), as well as the concise summary in \cite{LA} or the more extensive discussion in \cite{LA3} (in the context of two-field cosmological models). As a first step, note that the analogue of (4.1) in \cite{ADGW}, i.e. the reduced Lagrangian for the set of generalized coordinates $\{a, \varphi , \theta \}$, now is:
\be \label{L_cl_m}
{\cal L} \,= \,- 3 a \dot{a}^2 + \frac{a^3 \dot{\varphi}^2}{2} + \frac{a^3 f (\varphi) \,\dot{\theta}^2}{2} - a^3 V(\varphi,\theta) - \rho_{m_0} \,\,\, .
\ee
The next steps are to perform in (\ref{L_cl_m}) a coordinate transformation $(a, \varphi , \theta) \rightarrow (u,v,w)$\,, where $q^i \equiv \{u,v,w\}$ are generalized coordinates adapted to the above mentioned hidden symmetry, and then to solve the resulting simplified Euler-Lagrange equations, together with the Hamiltonian constraint $E_{\cal L} \equiv \frac{\pd {\cal L}}{\pd \dot{q}^i} \dot{q}^i - {\cal L} = 0$\,. The latter is necessary to ensure that the Friedman equation (\ref{EinstEq_m}) is solved too, and not just the scalar field ones (\ref{ScalarEoMs}). Note that inside the expression for $E_{\cal L}$\,, resulting from (\ref{L_cl_m}), both the $f$ and $V$ terms have a positive sign, unlike the relative minus between them in (\ref{L_cl_m}). Hence, even assuming the ansatz (\ref{th_d_om}), it is not possible to absorb $f$ in a redefinition of $V$, just as it is not in the original system given by (\ref{ScalarEoMs}) and (\ref{EinstEq_m}).

Since the only matter contribution in (\ref{L_cl_m}) is an additive constant, the change of variables, that simplifies greatly the equations of motion, is the same as in \cite{ADGW}, namely: 
\bea \label{Ch_var}
a (t) &=& \left[ u^2 - \left( v^2 + w^2 \right) \right]^{1/3} \,\,\,\, , \nn \\
\varphi (t) &=& \sqrt{\frac{8}{3}} \,{\rm arccoth} \!\left( \sqrt{\frac{u^2}{v^2 + w^2}} \,\,\right) \,\,\,\, , \nn \\
\theta (t) &=& \theta_0 + {\rm arccot} \!\left( \frac{v}{w} \right) \,\,\,\, , \,\,\,\, \theta_0 = const \quad .
\eea
The resulting Euler-Lagrange equations for the new variables $u$, $v$ and $w$ have the general solutions: 
\bea \label{Sols_uvw}
u (t) &=& C_1^u \sinh (\tilde{\kappa} t) + C_0^u \cosh (\tilde{\kappa} t) \quad , \nn \\
v (t) &=& C_1^v \sin (\omega t) + C_0^v \cos(\omega t) \quad , \nn \\
w (t) &=& C_1^w \sin (\omega t) + C_0^w \cos(\omega t) \quad ,
\eea
where $C_{0,1}^{u,v,w} = const$ and
\be \label{tkappa}
\tilde{\kappa} \equiv \frac{1}{2} \sqrt{3 C_V - 4 \omega^2} \,\,\, .
\ee
Note that the parameter $\tilde{\kappa}$\,, defined in (\ref{tkappa}), is related to the parameter $\kappa$ introduced in (\ref{k_def}) as $\tilde{\kappa} = \omega \sqrt{\frac{3}{4} \kappa -1}$\,.

The essential difference from \cite{ADGW} is in the Hamiltonian constraint. To explain that, let us use (\ref{Ch_var}) inside (\ref{L_cl_m}), in order to compute the resulting $E_{\cal L}$\,. We find:
\be \label{E_L}
E_{\cal L} = - \frac{4}{3} \dot{u}^2 + \frac{4}{3} \dot{v}^2 + \frac{4}{3} \dot{w}^2 + \frac{4}{3} \tilde{\kappa}^2 u^2 + \frac{4}{3} \omega^2 v^2 + \frac{4}{3} \omega^2 w^2 + \rho_{m_0} \,\,\, .
\ee
Substituting (\ref{Sols_uvw}) in (\ref{E_L}) and imposing $E_{{\cal L}} = 0$\,, we obtain:
\be \label{Constr}
\tilde{\kappa}^2 \!\left[ (C_0^u)^2 - (C_1^u)^2 \right] + \omega^2 \left[ (C_0^v)^2 + (C_1^v)^2 + (C_0^w)^2 + (C_1^w)^2 \right] + \frac{3 \rho_{m_0}}{4} = 0 \,\,\, .
\ee
Setting $C_1^v,C_0^w = 0$ and $C_0^v = C_1^w \equiv C_w$ inside (\ref{Sols_uvw}), as well as in (\ref{Constr}), gives precisely the $\dot{\theta} = \omega$ solutions of \cite{ADGW} (recalled in (\ref{Sol_om}) here), but now with matter included via the constraint (\ref{Constr}). Alternatively, one could keep all integration constants in (\ref{Sols_uvw}) arbitrary, modulo the relation (\ref{Constr}), in which case one obtains solutions with varying $\dot{\theta}$\,. It would be interesting to explore the properties of these more general solutions in future work.

It is easy to verify that indeed, with $f$ and $V$ as in (\ref{fs}) and (\ref{Vs}) respectively, the substitution of (\ref{Sols_uvw}) in (\ref{Ch_var}) gives solutions to the equations of motion (\ref{ScalarEoMs}) and (\ref{EinstEq_m}), upon using (\ref{Constr}). Of course, to obtain a realistic cosmological model, which can describe the transition between the epochs of matter domination and dark energy, one has to choose a suitable part of parameter space. Namely, the (so far arbitrary) overall scale of the dark energy contribution has to be taken such that, at some initial moment, that contribution is negligible compared to $\rho_m$\,. Then, (\ref{EinstEq_m}) would reduce to the standard matter-domination relation:
\be \label{H_rho_m}
3 H^2 \approx \rho_m \,\,\, .
\ee 
Note that during matter domination the dark energy scalars are effectively frozen, as mentioned in Section \ref{Prelim}, and hence $f$ and $V$ are effectively constant. The latter functions will start varying appreciably, only when the decreasing $\rho_m$ becomes comparable to the energy density of the dark sector and thus $\varphi$ starts rolling (see also \cite{ASSV}). In addition, note that at late times $V$ in (\ref{Vs}) tends to a constant again, as $\varphi$ tends to zero.

\subsection{On the $\sigma_8$ tension} \label{S8_ten}

As discussed above, the inclusion of matter allows us, in principle, to study the transition between the epochs of matter domination and dark energy domination, as well as to address the cosmological $\sigma_8$ parameter. However, in view of the large parameter space of our solutions, it seems that such a study would be technically involved and, probably, would require numerical methods. Furthermore, since we are interested in the field-space region with small $\varphi$ (as we saw in Section \ref{BehaviorCs}), performing numerical computations may be rather subtle and demanding due to numerical instabilities.\footnote{Indeed, see for example \cite{YW,AW} where the numerical computations in a model of cosmic acceleration (albeit in the context of orbital inflation) were intentionally restricted to the part of field space with a radial coordinate $> 1$\,, precisely because of numerical instabilities for radial coordinate $<1$\,. \label{OrbInfl}} For all these reasons, we leave for the future a detailed investigation, within our model, of the transition from matter domination to the dark energy epoch.

Nevertheless, in view of the results of \cite{MT}, we can already argue that our exact solutions should lead to slight lowering of the cosmological $\sigma_8$ parameter, compared to the $\Lambda$CDM model. Indeed, our equation-of-state parameter $w_{DE}$ is very close to $-1$\,, as discussed in \cite{ADGW} and Appendix \ref{w_DE} here. In addition, our speed of sound is $c_s \approx 0.4$, as shown in Section \ref{BehaviorCs} here. Therefore, our model lies on the right-hand side of Figure 7 in \cite{MT}\footnote{That Figure illustrates the general dependence of $\sigma_8$ on the equation-of-state parameter $w_{DE}$ and the speed of sound $c_s$ of dark energy.}, just below the $\Lambda$CDM prediction.

A note of caution is due to the fact that the considerations of \cite{MT} (see also \cite{HJ}) assume, for simplicity, a constant equation of state parameter, while in our solutions $w_{DE}$ is a function of time. However, recall that our solutions reach their asymptotic regime exponentially fast. Hence it is reasonable to expect that a detailed numerical computation in our case would give a result (broadly) in agreement with \cite{MT}. Thus our model has the potential to resolve the $\sigma_8$ tension.

It should be noted that the precise magnitude of this tension is still subject to controversy. See, for instance, the recent analysis in \cite{finalPl} of the observational data from the Planck satellite, which claims that the $\sigma_8$ tension is smaller than previously thought. It would certainly be very interesting to explore how the $\sigma_8$-prediction of our model depends on the multi-dimensional parameter space, as well as to compare that prediction to the latest estimate obtained from the observational data. We hope to be able to report on that in the future.

\subsection{Hubble tension}

Let us now discuss the implications of our model for the Hubble tension. The latter is a discrepancy between the value of the Hubble parameter inferred from the CMB (i.e. the early Universe), and evolved to the present with the $\Lambda$CDM model, and its value extracted from observations of the late Universe. Many models in the literature aim to alleviate this tension by introducing new physics at early times, before the epoch of recombination. However, there is growing evidence that to resolve completely the Hubble tension, it is (also) necessary to modify the late Universe; see for instance \cite{HW,SV} and references therein.

So we turn to investigating whether, in our model of the late Universe, it is possible to obtain a present-day value of the Hubble parameter (i.e., Hubble constant) that is larger than in $\Lambda$CDM, which would alleviate the tension with the early Universe. To facilitate the comparison between the $\Lambda$CDM and our models, in this Section we will denote the Hubble parameter by $H_{\Lambda}$\,, in the case of a cosmological constant $\Lambda$\,, and by $H_{DE}$\,, in the dynamical dark energy case.

Then the analogue of the Friedman equation (\ref{EinstEq_m}), for the case of a cosmological constant, is:
\be \label{H_L}
3 H^2_{\Lambda} = \Lambda + \rho_m \,\,\, ,
\ee
where the vacuum energy density $\Lambda = const$\,. For convenience, let us also rewrite (\ref{EinstEq_m}) in the notation of this Section:
\be \label{H_DE}
3 H^2_{DE} = \frac{1}{2} \!\left( \dot{\varphi}^2 + f \dot{\theta}^2 \right) + V (\varphi) + \rho_m \,\,\, .
\ee
Note that at early times, i.e. during matter domination, the Hubble parameter is the same in both cases, since neglecting $\Lambda$ and the dark sector in (\ref{H_L}) and (\ref{H_DE}) gives:
\be \label{H_L_DE_in}
3 H_{\Lambda}^2 \approx \rho_m \qquad {\rm and} \qquad 3 H^2_{DE} \approx \rho_m \,\,\, .
\ee
Hence we need to show that our solutions, when starting (during matter domination) with the same value of $H (t)$ as in $\Lambda$CDM, can evolve with time in such a way, that in at least some part of parameter space we have: 
\be \label{Ht*}
H_{DE} (t_*) > H_{\Lambda} (t_*) \,\,\, ,
\ee
where  $t_*$ is the present.

To see whether this is possible, let us consider the dark energy contribution to the Hubble parameter, namely:
\be \label{E_DE}
E_{DE} \equiv \frac{1}{2} \!\left( \dot{\varphi}^2 + f \dot{\theta}^2 \right) + V (\varphi) \,\,\, .
\ee
The kinetic term in (\ref{E_DE}) is positive-definite and tends to zero with time, as shown in \cite{ADGW}. Hence, to ensure that $E_{DE} > \Lambda$ for some period of time, it is enough to have $V (\varphi) > \Lambda$ during that time. Now note that the potential (\ref{Vs}) can be rewritten as:
\be \label{V_k}
V (\varphi) = \omega^2 \left[ \kappa \,\cosh^2 \!\left( \frac{\sqrt{6}}{4} \,\varphi \right) - \frac{4}{3} \right] \, ,
\ee
in terms of the parameter introduced in (\ref{k_def}). The considerations of the previous Section constrain $\kappa$\,, but not $\omega$\,.\footnote{Note that $\omega$ cancels out of (\ref{cs_subl}) and (\ref{large_eta}) upon substituting $C_V = \kappa \,\omega^2$, in accordance with (\ref{k_def}). Hence, assuming small field values in line with Section \ref{BehaviorCs}, we are free to choose $\omega$ as needed in this Section, without affecting the previous considerations. Furthermore, clearly the overall scale of (\ref{E_DE}) at any moment depends also on the integration constants in the solutions for the dark energy scalars. So, in the rapid turn regime, we have plenty of freedom to ensure the initial condition (\ref{H_L_DE_in}).} So we can choose the arbitrary constant $\omega$ such that, at some initial moment of time $t_0$ during matter domination, we have: 
\be \label{VgrVL}
V(\varphi)|_{t=t_0} > \Lambda \, .
\ee
Since $\varphi (t)$ is monotonically decreasing \cite{ADGW}, the potential (\ref{V_k}) is also a monotonically decreasing function of time. So if we start from the initial condition (\ref{VgrVL}), then it is clear from (\ref{H_L}) and (\ref{H_DE}) that we will have $H_{DE} (t) > H_{\Lambda} (t)$ for a period of time beginning with $t_0$\,. Of course, the duration of that period, and whether it can last until $t_*$ (i.e. until the present), will depend on the specific choices of values for the integration constants in the solution. However, if take $\omega$ such that the minimum of $V(\varphi)$ is greater than $\Lambda$\,, i.e. such that $\omega^2 (\kappa - \frac{4}{3}) > \Lambda$\,, then the inequality
\be
H_{DE} (t) > H_{\Lambda} (t)
\ee
will be satisfied for any $t$ (including, of course, $t_*$).\footnote{Recall that we have assumed small $\varphi$\,, in which case there are no constraints on $\omega$ from the previous considerations. However, even for $\varphi$ of order $1$ or larger, it is not difficult to satisfy the constraint on $\omega$\,, which according to Appendix \ref{r_of_varphi} would arise from keeping $M_{eff}^2$ small, simultaneously with satisfying the inequality $\omega^2 (\kappa - \frac{4}{3}) > \Lambda$\,, due to $\Lambda <\!\!< 1$ in our units in which $M_p=1$\,.} To summarize, by choosing suitably the value of the arbitrary constant $\omega$ in our solutions, we can easily ensure that the inequality (\ref{Ht*}) is satisfied. 

Note that, as discussed above, to ensure (\ref{Ht*}) we need to choose initial conditions satisfying:
\be \label{Vineq}
\left[ \frac{1}{2} \!\left( \dot{\varphi}^2 + f \dot{\theta}^2 \right) + V (\varphi) \right] \bigg|_{t=t_0} > \,\Lambda \,\,\, .
\ee
Since for our solutions the function $E_{DE} (t)$ decreases with time, (\ref{Vineq}) implies the following. Starting, during matter domination, with $\rho_m (t)$ much greater than either of $E_{DE} (t)$ and $\Lambda$\,, the subsequent evolution, which leads to the present epoch of acceleration, will take different amounts of time in the two different models. Namely, the equality $\rho_m (t) = E_{DE} (t)$ will be reached earlier in time than the equality $\rho_m (t) = \Lambda$\,. In other words, the transitioning to the dark energy epoch will begin earlier in our model, than in $\Lambda$CDM, for solutions that alleviate the Hubble tension.  

It should be noted that the above arguments are rather general and, in particular, do not depend on the specific form of $\varphi(t)$\,. Neither is it crucial to have multiple scalar fields, as opposed to a single scalar. The most important property of the model is that the kinetic and potential energies of the dark energy field(s) decrease with time.\footnote{In contrast, the claim of \cite{BCHCSJY} is based on a particular single-field dark energy (quintessence) model, which includes a specific interaction between dark energy and dark matter.} However, it should be kept in mind that satisfying (\ref{Ht*}) is only a necessary condition for alleviating the Hubble tension. One should also investigate whether solutions with this property are consistent with the observational constraints on other cosmological parameters.\footnote{A particularly stringent restriction is due to the comparison of \cite{A4B2etal} and \cite{Reissetal} with \cite{PZJ,eBOSS}. We will discuss in more detail its implications in Section \ref{Discussion}.} It would certainly require a much more thorough (and, possibly, numerical) study of our class of exact solutions, in order to answer that question.\footnote{Regarding the subtleties that such a numerical computation would face, see the discussion in the beginning of Subsection \ref{S8_ten} and, specifically, footnote \ref{OrbInfl}.} So we leave its investigation for the future.

\section{Discussion} \label{Discussion}

We studied the speed of sound $c_s$ of dark energy perturbations around the background solution of \cite{ADGW}. This is a multifield model of dark energy, which relies on rapid-turning trajectories in field space. Unlike single-field models that always have $c_s = 1$\,, the multifield case allows for the possibility that $c_s < 1$ (even when the equation-of-state parameter is arbitrarily close to $-1$). Such reduced sound speed is of great interest, as it can lead to qualitatively different effects. We showed that the rapid turn regime of our solutions is, indeed, correlated with a reduced speed of sound. Furthermore, we proved that the latter is a monotonically decreasing function of time, everywhere in our parameter space. We also found that $c_s^2 (t)$ has a positive-definite lower bound. This is unlike the function $c_s^2 (t)$ in the multifield dark energy model of \cite{ASSV,EASV}. Although that model has an equation-of-state parameter $w_{DE} \approx -1$, similarly to ours (recall Appendix \ref{w_DE}), their $c_s^2 (t)$ can be arbitrarily small and, in some parts of their parameter space, it can become negative with time, which indicates a gradient instability. Our model, by contrast, does not have such instabilities anywhere in its parameter space. In addition, we argued that the positive-definite lower bound of our $c_s$ leads to a characteristic clustering scale of order $6$ Gpc. Although detecting the effects of dark energy clustering on such large scales is beyond the reach of current observations, it might become accessible to future ones.

We also found a class of exact solutions to the background equations of motion, which incorporates matter in the dark energy solutions of \cite{ADGW}. This enabled us to discuss the implications of our class of models for the $\sigma_8$ and Hubble tensions. In particular, on the basis of our results for the equation-of-state parameter $w_{DE}$ and sound speed $c_s$\,, we argued that the $\sigma_8$ tension should be alleviated, regardless of the choices of parameter values in the background solution. Alleviating the Hubble tension, on the other hand, does involve restrictions on parameter space. Satisfying those restrictions is related to beginning earlier, than in $\Lambda$CDM, the transition from matter domination to the current epoch of accelerated expansion. It has to be noted that, unlike the detailed investigation of $c_s$ in the bulk of the paper, the discussion of the two cosmological tensions in Section \ref{InclM} is rather preliminary. Nevertheless it serves as a strong motivation for further studies of multifield dark energy, given the well-known in the literature difficulty in alleviating one tension without worsening the other.

It should be stressed that we have studied a model of the late Universe. Hence we cannot address issues related to the sound horizon $r_d$ of the photon-baryon fluid at the epoch of baryon drag. Models, which introduce new physics in the early Universe to resolve the Hubble tension, aim to lower $r_d$ in order to be consistent with the observational constraints \cite{PZJ,eBOSS} on its product with the Hubble constant. However, relying on such early-time resolutions alone leads to problems regarding the observational constraints on the matter density and the $\sigma_8$ parameter \cite{SV}. This is not the only indication that late-time physics, different from $\Lambda$CDM, is necessary. A growing amount of evidence for that is provided by the observed ``descending trends'' in the determination of the Hubble constant from a variety of low-redshift data \cite{DescendingT,DescendingT2,DescendingT3,DescendingT4,DescendingT5,DescendingT6,DescendingT7,DescendingT8,DescendingT9}. Namely, these works show that, assuming the $\Lambda$CDM model, the inferred value of the Hubble constant decreases with increasing redshift of the observations. This implies that there is a need for deviation from $\Lambda$CDM in the late Universe. Whether our model can provide the necessary modification is, undoubtedly, very interesting to explore.\footnote{For other modifications of the late Universe, which aim to relieve the Hubble tension, see \cite{SPYY,GSS,MP,HW2}.}

It is also worth investigating the consequences of our class of dark energy models for the large-scale structure of the Universe at late times. For that purpose, one should study in much more detail the equations of motion of our dark energy fluctuations, including interaction with matter perturbations and gravitational backreaction. This is likely to involve demanding numerical computations, as well as the use of the explicit background solutions for $H(t)$ and $\varphi (t)$\,. Here we managed to avoid using the explicit form of these solutions. Thus, our conclusions depended only implicitly, through the initial condition $\varphi_0$\,, on all integration constants, except those in the potential. However, extracting phenomenological predictions, regarding the effects of dark energy clustering on the large-scale clustering of matter, most likely will require identifying a suitable part of the full parameter space of the model. We hope to make progress on this problem in the future.

\section*{Acknowledgements}

We would like to thank A. Buchel, C. Lazaroiu, S. Parameswaran, P. Suranyi and R. von Unge for interesting discussions. L.A. also thanks the Stony Brook Simons Workshop in Mathematics and Physics for hospitality during the completion of this work. L.A. has received partial support from the Bulgarian NSF grant KP-06-N68/3. The research of L.C.R.W. is partially supported by the US. Department of Energy grant DE-SC1019775. J.D. is funded by a University of Cincinnati undergraduate research fellowship.

\appendix

\section{Properties of $r (\varphi)$ and $M_{eff}^2 (\varphi)$} \label{r_of_varphi}
\setcounter{equation}{0}

In our class of dark energy solutions, the function $\varphi (t)$ is monotonically decreasing and with time tends to $\varphi = 0$\,. Hence whether or not large field values are probed by a specific solution is determined by the initial value of $\varphi$\,. So let us investigate the ratio $r$\,, given by (\ref{r_def}), and the mass $M_{eff}^2$\,, determined from (\ref{MTMeff}), as functions of $\varphi$\,, in order to see in what part of field space the approximation (\ref{cs_2}) is reliable.

We begin by studying the behavior of $r(\varphi)$\,. To find this function explicitly, we substitute (\ref{th_d_om})-(\ref{Vs}) and (\ref{ph_d_ph}) inside (\ref{Om})-(\ref{VTT}). Using those results in (\ref{MTMeff}) and then in (\ref{r_def}), we obtain:
\be \label{r_DN}
r (\varphi) \,= \,\frac{4 M_1 M_2 M_3}{r_D} \,\,\, ,
\ee
where
\bea \label{M123}
M_1 &=& (3 \kappa + 4)^2 \cosh^6 \!\hat{\varphi} - 4 (3 \kappa + 4) \cosh^4 \!\hat{\varphi} - 8(3 \kappa + 2) \cosh^2 \!\hat{\varphi} + 16 \,\,\, , \nn \\
M_2 &=& 2 (3\kappa + 4) \cosh^4 \!\hat{\varphi} - (3\kappa + 16) \cosh^2 \!\hat{\varphi} + 8 \,\,\, , \nn \\
M_3 &=& (3\kappa + 4) \cosh^2 \!\hat{\varphi} - 4 
\eea
and
\be
r_D = \left[ 3 (3 \kappa + 4)^2 \cosh^6 \!\hat{\varphi} - (9 \kappa^2 + 96 \kappa + 112) \cosh^4 \!\hat{\varphi} + 20 (3 \kappa + 4 ) \cosh^2 \!\hat{\varphi} - 16 \right]^2 .
\ee
Here, for convenience, we have used again the notation (\ref{phi_hat}). Note that in the limit $\varphi \rightarrow \infty$ the parameter $\kappa$ cancels out of (\ref{r_DN}) and we find:
\be
r (\varphi) \rightarrow \frac{8}{9} \,\,\, .
\ee
On the other hand, for $\varphi \rightarrow 0$ the expression (\ref{r_DN}) reduces to (\ref{r_as}).
\begin{figure}[t]
\begin{center}
\hspace*{-0.3cm}
\includegraphics[scale=0.34]{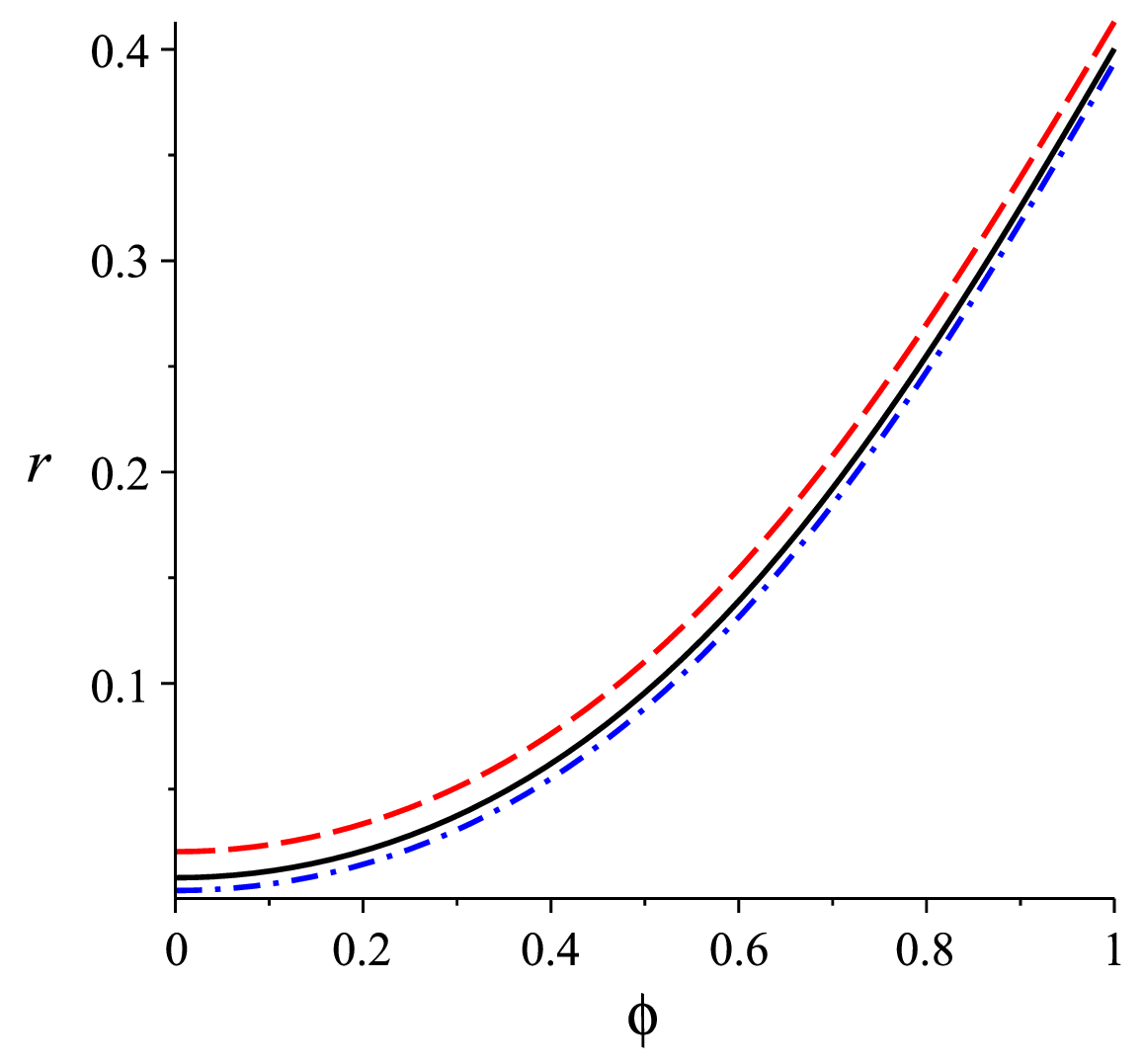}
\hspace*{0.7cm}
\includegraphics[scale=0.34]{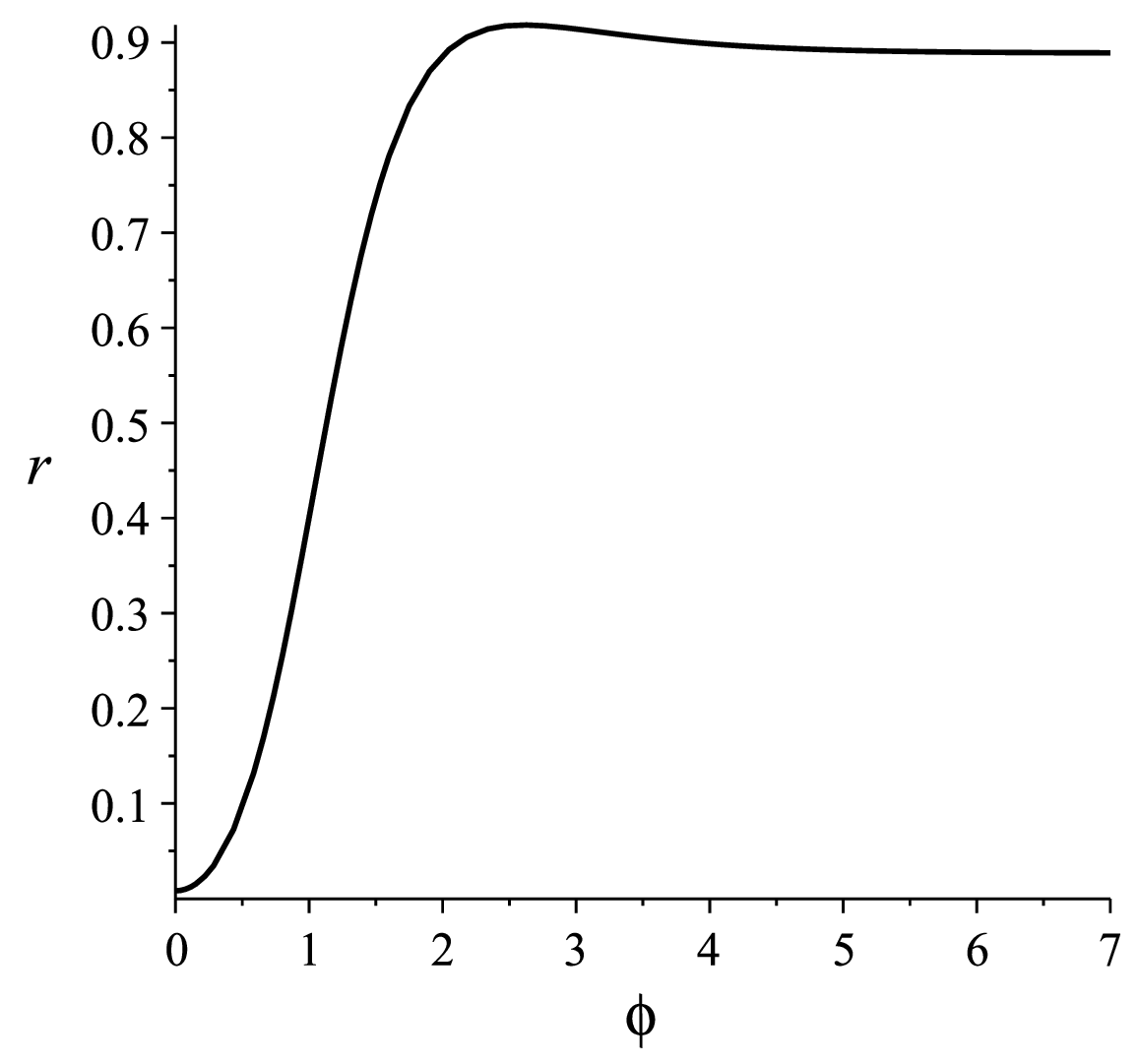}
\end{center}
\vspace{-0.7cm}
\caption{{\small Plots of the function $r (\varphi)$\,, given by (\ref{r_DN}), for $\kappa = 1.35$ (blue dash-dotted curve), $\kappa = 1.4$ (black solid curve) and $\kappa = 1.5$ (red dashed curve). The three graphs are indistinguishable on the right plot.}}
\label{r-ph}
\vspace{0.1cm}
\end{figure}

In fact, it turns out that $r \approx 0.89$ is a good approximation numerically for any $\varphi \gtrsim 2$ and also that, in the relevant part of parameter space, the dependence of $r$ on $\kappa$ is very weak. We exhibit the entire behavior of the function $r (\varphi)$ in Figure \ref{r-ph}. On the left of the Figure, we have plotted (\ref{r_DN}) for three values of $\kappa$\,, in order to illustrate how weak the dependence on that parameter is in the rapid-turning regime; in that regard, note that $\eta_{\perp}^2 \lesssim 20$ for $\kappa \ge 1.5$\,, while $\eta_{\perp}^2 \sim {\cal O} (100)$ for $\kappa = 1.35$\,, as is evident from Figure \ref{ParSpace}. On the right of Figure \ref{r-ph}, we have exhibited the entire shape of the function $r (\varphi)$\,. On this plot the three graphs, corresponding to the three values of $\kappa$\,, are indistinguishable from each other. From Figure \ref{r-ph} we can see that, in general, $r$ is small for any $\varphi < 1$\,, but the inequality $r <\!\!< 1$ is really well-satisfied for $\varphi \lesssim 0.5$\,. On the other hand, for large field values $r \sim {\cal O} (1)$, although it is never $>\!\!> 1$\,. Hence, if the initial condition $\varphi_0$ is in the region of small field values, the approximation (\ref{Ineq}) is well-satisfied for the entire duration of the evolution. Whereas if $\varphi_0$ is not small, then at early times the speed of sound may deviate to some degree from (\ref{cs_2}), although with time it will approach fast the predictions of that formula.

\begin{figure}[t]
\begin{center}
\hspace*{-0.4cm}
\includegraphics[scale=0.35]{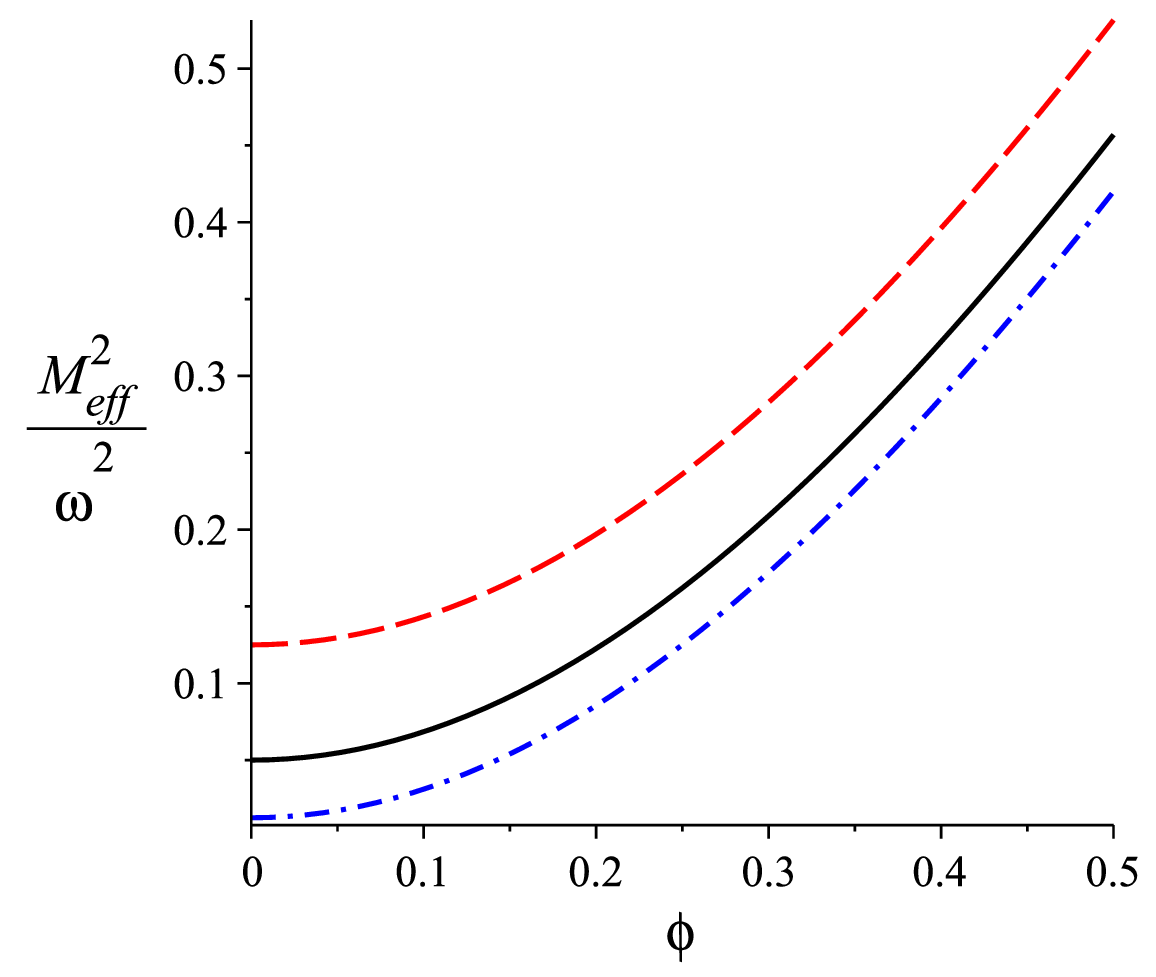}
\hspace*{0.6cm}
\includegraphics[scale=0.362]{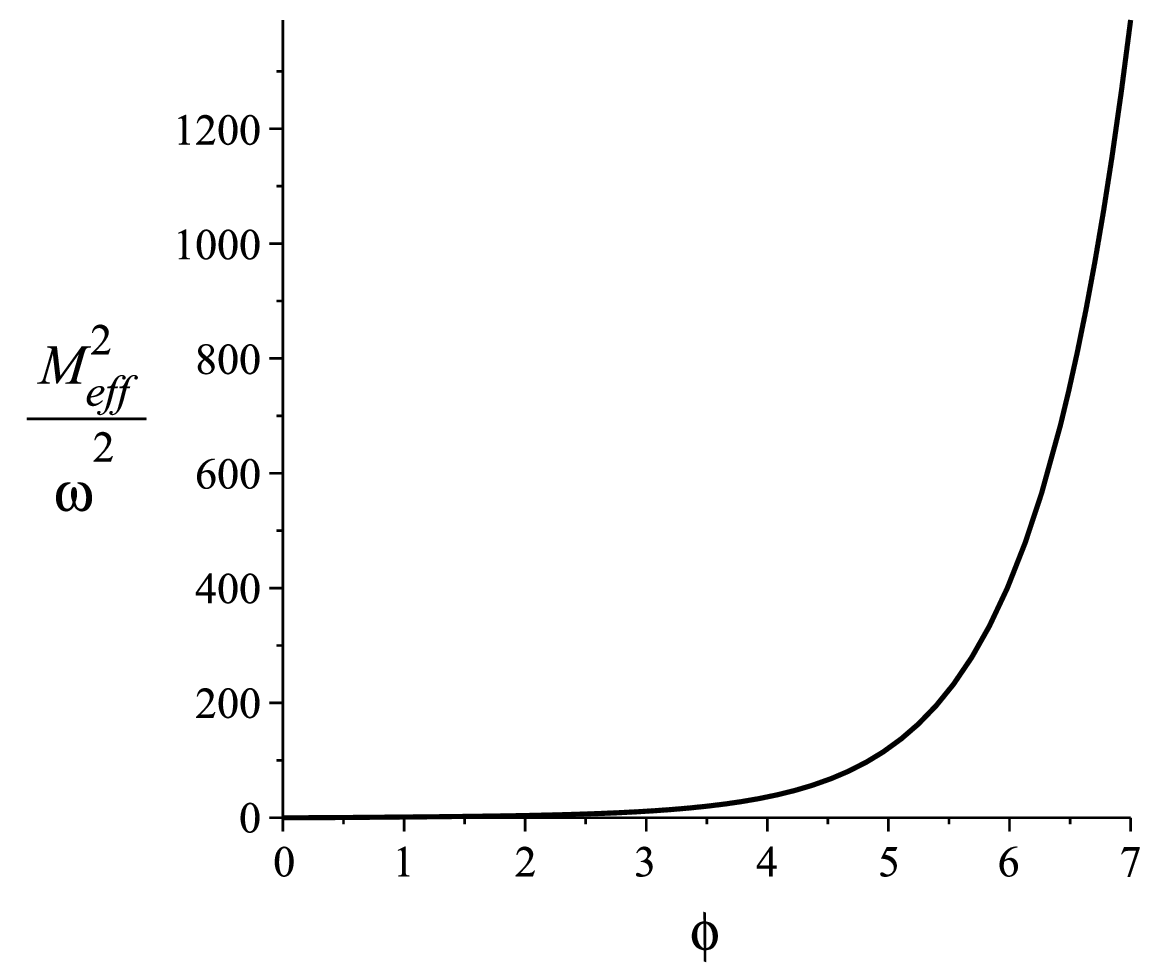}
\end{center}
\vspace{-0.7cm}
\caption{{\small Plots of $M_{eff}^2/\omega^2$ as a function of $\varphi$\,, according to (\ref{Meff_ph}), for $\kappa = 1.35$ (blue dash-dotted curve), $\kappa = 1.4$ (black solid curve) and $\kappa = 1.5$ (red dashed curve). On the right-side plot, the three graphs coincide.}}
\label{Meff_sq-ph}
\vspace{0.1cm}
\end{figure}
Now let us turn to investigating the function $M_{eff}^2 (\varphi)$\,. In the case of our background solutions, (\ref{MTMeff}) gives:
\be \label{Meff_ph}
M_{eff}^2 (\varphi) \,= \,\frac{3}{4} \,\omega^2 \frac{\kappa \,M_1}{\left[(3 \kappa + 4) \cosh^2 \!\hat{\varphi} - 4\right]^2} \,\,\, ,
\ee
where $M_1$ is the same as in (\ref{M123}). For $\varphi <\!\!< 1$\,, this gives $M_{eff}^2 \approx \omega^2 \!\left(\frac{3}{4} \kappa - 1 \right)$\,, as discussed in Section \ref{BehaviorCs}, which leads to small $M_{eff}^2$ in the rapid-turning regime. However, for $\varphi >\!\!> 1$\,, (\ref{Meff_ph}) implies that $M_{eff}^2 \approx \frac{3}{16} \omega^2 \kappa \,e^{2 \hat{\varphi}}$\,. We have illustrated the behavior of the ratio $M_{eff}^2/\omega^2$ as a function of $\varphi$ in Figure \ref{Meff_sq-ph}, for the same three choices of $\kappa$ as on Figure \ref{r-ph}. On the left side of Figure \ref{Meff_sq-ph}, we have focused on the small-field region, restricting to $\varphi \lesssim 0.5$ in view of the above discussion of the behavior of $r (\varphi)$\,. On the right side of the Figure, we exhibit the rapid growth of $M_{eff}^2 / \omega^2$ for $\varphi > 1$\,; as before, on this plot the three graphs, corresponding to the three values of $\kappa$\,, are indistinguishable from each other. Of course, even if the initial condition $\varphi_0$ is in the large-field region, one can still ensure that the initial value of $M_{eff}^2$ is as small as desired, by choosing suitably the arbitrary constant $\omega$\,. However, the behavior of the function $M_{eff}^2 / \omega^2$ with increasing $\varphi$ indicates, again, that the small-field region of field space is the natural domain of validity of our model.

\section{Behavior of $w_{DE} (N)$} \label{w_DE}
\setcounter{equation}{0}

In this Appendix we investigate the dependence of the equation-of-state parameter $w_{DE}$\,, of our dark energy solutions, on the number of e-folds $N$ of accelerated expansion. This will allow us to estimate how fast the solutions approach their asymptotic regime, in which $w_{DE} \approx -1$\,.

Using (\ref{Hf}), we find for the number of e-folds:
\be
N = \int H dt = - \int \frac{f'}{3 f} d \varphi = - \frac{1}{3} \ln f + const \,\,\, .
\ee
To fix the integration constant, let us require that $N=0$ at some initial moment of time $t_0$\,, namely:
\be \label{Nf}
N = - \frac{1}{3} \ln \frac{f}{f_0} \quad , \quad {\rm where} \quad f_0 \equiv f(\varphi)|_{t=t_0} \,\,\, .
\ee
Note that, since in the solutions of \cite{ADGW} the function $\varphi (t)$ is monotonically decreasing, the function $f(\varphi)$ in (\ref{fs}) is also monotonically decreasing. Hence the number of e-folds, determined by (\ref{Nf}), is always $N > 0$ for any $t > t_0$\,, as should be the case. Furthermore, there is no upper limit for $N$\,, i.e. $N \rightarrow \infty$ when $t \rightarrow \infty$\,. For the subsequent considerations, it will be useful to invert (\ref{Nf}), in order to express $f$ as a function of $N$:
\be \label{fN}
f = f_0 e^{-3N} \,\,\, .
\ee

Now let us consider the equation-of-state parameter of the dark energy solutions found in \cite{ADGW}:
\be \label{wDE}
w_{DE} (t) = \frac{\frac{1}{2}\left( \dot{\varphi}^2 + f \dot{\theta}^2 \right) -V}{\frac{1}{2}\left( \dot{\varphi}^2 + f \dot{\theta}^2 \right) +V} \,\,\, .
\ee
Substituting (\ref{th_d_om}) and (\ref{ph_d}) in (\ref{wDE}), we obtain:
\be \label{w_DE_fV}
w_{DE} (\varphi) = \frac{2 V}{\frac{f'^2}{3f^2} (f \omega^2 + 2V)} + \frac{f \omega^2 - 2 V}{f \omega^2 + 2 V} \,\,\, .
\ee
To convert this into a function of the number of e-folds $N$\,, note that we can write the derivative of the function (\ref{fs}), namely $f' (\varphi) = 2 \sqrt{\frac{8}{3}} \sinh \!\left( \sqrt{\frac{3}{8}} \,\varphi \right) \cosh \!\left( \sqrt{\frac{3}{8}} \,\varphi \right)$\,, as:
\be \label{fp_f}
f' = 2 \,\sqrt{f \left( 1 + \frac{3}{8} f \right)} \,\,\, ,
\ee
while the potential (\ref{Vs}) as:
\be \label{V_f}
V = C_V \left( 1 + \frac{3}{8} f \right) - \frac{4}{3} \omega^2 \,\,\, .
\ee
Hence using (\ref{fp_f}) and (\ref{V_f}), together with (\ref{k_def}), inside (\ref{w_DE_fV}) gives:
\be \label{wDEf}
w_{DE} (\varphi) = \frac{3}{2} \frac{ f \left( \kappa-\frac{4}{3} + \frac{3}{8} \kappa f \right) }{\left( 1+\frac{3}{8} f \right) \left[ \,2 \!\left( \kappa - \frac{4}{3} \right) + \left( 1+\frac{3}{4} \kappa \right) \!f \right]} + \frac{\left( 1-\frac{3}{4} \kappa \right) f - 2 \left( \kappa - \frac{4}{3} \right)}{\left( 1+\frac{3}{4} \kappa \right) f + 2 \left( \kappa - \frac{4}{3} \right)} \,\, .
\ee
Finally, substituting (\ref{fN}) in (\ref{wDEf}), we obtain:
\be \label{wDEN}
w_{DE} (N) = \frac{3}{2} \frac{ f_0 e^{-3N} \!\left( \kappa-\frac{4}{3} + \frac{3}{8} \kappa f_0 e^{-3N} \right) }{\left( 1+\frac{3}{8} f_0 e^{-3N} \right) \!\left[ \,2 \!\left( \kappa - \frac{4}{3} \right) + \left( 1+\frac{3}{4} \kappa \right) \!f_0 e^{-3N} \right]} + \frac{\left( 1-\frac{3}{4} \kappa \right) \!f_0 e^{-3N} - 2 \left( \kappa - \frac{4}{3} \right)}{\left( 1+\frac{3}{4} \kappa \right) \!f_0 e^{-3N} + 2 \left( \kappa - \frac{4}{3} \right)} \, .
\ee
Note that numerically $e^{-3} <\!\!< 1$\,. So it is illuminating to expand the expression (\ref{wDEN}) in small $e^{-3N}$:
\be
w_{DE} (N) = - 1 + \frac{3}{4} f_0 e^{-3N} + {\cal O} (e^{-6N}) \,\,\, .
\ee
This makes it evident that the equation of state parameter $w_{DE}$ approaches $-1$ very fast as the number of e-folds $N$ increases. In fact, already the first e-fold leads to $w_{DE} \approx -1$\,, when $f_0 \sim {\cal O} (1)$ as is the case of interest for us. Indeed, in view of 
Appendix \ref{r_of_varphi}, 
we are interested in initial conditions that are $\varphi_0 < 1$ or even $\varphi_0 \lesssim 0.5$\,. To be more explicit, note that $f|_{\varphi_0 = 1} = 1.13$\,, while $f|_{\varphi_0 = 0.5} = 0.26$\,.


\begin{thebibliography}{100}

\bibitem{A4B2etal}
N. Aghanim et al., {\em Planck 2018 results}, Astron. Astrophys. 641 (2020) A6, [Erratum: Astron. Astrophys. 652 (2021) C4], arXiv:1807.06209 [astro-ph.CO].

\bibitem{Reissetal}
A. G. Riess et al., {\em A Comprehensive Measurement of the Local Value of the Hubble Constant with 1 km/s/Mpc Uncertainty from the Hubble Space Telescope and the SH0ES Team}, Astrophys. J. Lett. 934 (2022) L7, arXiv:2112.04510 [astro-ph.CO].

\bibitem{Heymetal}
C. Heymans et al., {\em KiDS-1000 Cosmology: Multi-probe weak gravitational lensing and spectroscopic galaxy clustering constraints}, Astron. Astrophys. 646 (2021) A140, arXiv:2007.15632 [astro-ph.CO].

\bibitem{Valetal}
E. Di Valentino et al., {\em Cosmology Intertwined III: f$\sigma_8$ and $S_8$}, Astropart. Phys. 131 (2021) 102604, arXiv:2008.11285 [astro-ph.CO]. 

\bibitem{NV}
R. C. Nunes and S. Vagnozzi, {\em Arbitrating the $S_8$ discrepancy with growth rate measurements from Redshift-Space Distortions}, Mon. Not. Roy. Astron. Soc. 505, 5427 (2021), arXiv:2106.01208 [astro-ph.CO].

\bibitem{OOSV}
G. Obied, H. Ooguri, L. Spodyneiko and C. Vafa, {\em De Sitter Space and the Swampland}, arxiv:1806.08362 [hep-th].

\bibitem{GK}
S. Garg and C. Krishnan, {\em Bounds on Slow Roll and the de Sitter Swampland}, JHEP 11 (2019) 075,  arXiv:1807.05193 [hep-th].

\bibitem{OPSV}
H. Ooguri, E. Palti, G. Shiu, C. Vafa, {\em Distance and de Sitter Conjectures on the Swampland}, Phys. Lett. B 788 (2019) 180, arXiv:1810.05506 [hep-th].

\bibitem{AP}
A. Achucarro and G. Palma, {\em The string swampland constraints require multi-field inflation}, JCAP 02 (2019) 041, arXiv:1807.04390 [hep-th].

\bibitem{BPR}
R. Bravo, G. A. Palma, S. Riquelme, {\em A Tip for Landscape Riders: Multi-Field Inflation Can Fulfill the Swampland Distance Conjecture}, JCAP 02 (2020) 004, 1906.05772 [hep-th].

\bibitem{AB} A. Brown, {\em Hyperinflation}, Phys. Rev. Lett. 121 (2018)
251601, arXiv:1705.03023 [hep-th].

\bibitem{SM}
S. Mizuno, S. Mukohyama, {\em Primordial perturbations from inflation with a hyperbolic field-space}, Phys. Rev. D 96 (2017) 103533, arXiv:1707.05125 [hep-th].

\bibitem{CRS} P. Christodoulidis, D. Roest, E. Sfakianakis, {\em Angular inflation in multi-field $\alpha$-attractors}, JCAP 11 (2019) 002, arXiv:1803.09841 [hep-th].

\bibitem{GSRPR} S. Garcia-Saenz, S. Renaux-Petel, J. Ronayne, {\em Primordial fluctuations and non-Gaussianities in sidetracked inflation}, JCAP 1807 (2018) 057, arXiv:1804.11279 [astro-ph.CO].

\bibitem{ACIPWW} A. Achucarro, E. Copeland, O. Iarygina, G. Palma, D.G. Wang, Y. Welling, {\em Shift-Symmetric Orbital Inflation:~single field or multi-field?}, Phys. Rev. D 102 (2020) 021302, arXiv:1901.03657 [astro-ph.CO].

\bibitem{TB} T. Bjorkmo, {\em The rapid-turn inflationary attractor}, Phys. Rev. Lett. 122 (2019) 251301, arXiv:1902.10529 [hep-th].

\bibitem{BM} T. Bjorkmo, M. C. D. Marsh, {\em Hyperinflation generalised:  from its attractor mechanism to its tension with the `swampland conditions'}, JHEP 04 (2019) 172, arXiv:1901.08603 [hep-th].

\bibitem{PSZ} G. A. Palma, S. Sypsas, C. Zenteno, {\em Seeding primordial black holes in multi-field inflation}, Phys. Rev. Lett. 125 (2020) 121301, arXiv:2004.06106 [astro-ph.CO].

\bibitem{FRPRW} J. Fumagalli, S. Renaux-Petel, J. W. Ronayne, L. T. Witkowski, {\em Turning in the landscape:~a new mechanism for generating Primordial Black Holes}, Phys. Lett. B 841 (2023) 137921, arXiv:2004.08369 [hep-th].

\bibitem{APR}
V. Aragam, S. Paban, R. Rosati, {\em The Multi-Field, Rapid-Turn Inflationary Solution}, JHEP 03 (2021) 009, arXiv:2010.15933 [hep-th].

\bibitem{LA}
L. Anguelova, {\em On Primordial Black Holes from Rapid Turns in Two-field Models}, JCAP 06 (2021) 004, arXiv:2012.03705 [hep-th].

\bibitem{LA2_pbh} L. Anguelova, {\em Primordial Black Hole Generation in a 
Two-field Inflationary Model}, Springer Proc. Math. Stat. 396 (2022) 193,  
arXiv:2112.07614 [hep-th].

\bibitem{CR}
P Christodoulidis, R. Rosati, {\em (Slow-)Twisting inflationary attractors}, JCAP 09 (2023) 034, arXiv:2210.14900 [hep-th].

\bibitem{IMS}
O. Iarygina, M.C.D. Marsh, G. Salinas, {\em Non-Gaussianity in rapid-turn multi-field inflation}, arXiv:2303.14156 [astro-ph.CO].

\bibitem{ChG}
P. Christodoulidis, J.-O. Gong, {\em Enhanced power spectra from multi-field inflation}, arXiv:2311.04090 [hep-th].

\bibitem{CDP}
M. Cicoli, G. Dibitetto, F. Pedro, {\em New accelerating solutions in late-time cosmology}, Phys. Rev. D 101 (2020) 10, 103524, arXiv:2002.02695 [gr-qc].

\bibitem{CDP2}
M. Cicoli, G. Dibitetto, F. Pedro, {\em Out of the Swampland with Multifield Quintessence?}, JHEP 10 (2020) 035, arXiv:2007.11011 [hep-th].

\bibitem{ASSV}
Y. Akrami, M. Sasaki, A. Solomon, V. Vardanyan, {\em Multi-field dark energy: cosmic acceleration on a steep potential}, Phys. Lett. B819 (2021) 136427, 	arXiv:2008.13660 [astro-ph.CO].

\bibitem{EASV}
J. Eskilt, Y. Akrami, A. Solomon, V. Vardanyan, {\em Cosmological dynamics of multifield dark energy}, Phys. Rev. D 106 (2022) 023512, arXiv:2201.08841 [astro-ph.CO].

\bibitem{ADGW}
L. Anguelova, J. Dumancic, R. Gass, L.C.R. Wijewardhana, {\em Dark Energy from Inspiraling in Field Space}, JCAP 03 (2022) 018, arXiv:2111.12136 [hep-th].

\bibitem{AAGP}
A. Achucarro, V. Atal, C. Germani, G. A. Palma, {\em Cumulative effects in inflation with ultra-light entropy modes}, JCAP 02 (2017) 013, arXiv:1607.08609 [astro-ph.CO].

\bibitem{LA2}
L. Anguelova, {\em Hidden Symmetries, Rapid Turns and Cosmic Acceleration}, PoS BPU11 (2023) 049, arXiv:2212.14127 [hep-th].

\bibitem{HS}
W. Hu, R. Scranton, {\em Measuring Dark Energy Clustering with CMB-Galaxy Correlations}, Phys. Rev. D70 (2004) 123002, arXiv:astro-ph/0408456.

\bibitem{ABL}
L. Anguelova, E.M. Babalic, C.I. Lazaroiu, {\em Two-field Cosmological $\alpha$-attractors with Noether Symmetry}, JHEP 1904 (2019) 148, arXiv:1809.10563 [hep-th].

\bibitem{LA3}
L. Anguelova, {\em Chapter 1: Hidden and Visible Symmetries in Two-Field Cosmological Models}, Peter Suranyi 87th Birthday Festschrift, World Scientific (2022) 1. 

\bibitem{YW}
Y. Welling, {\em A simple, exact, model of quasi-single field inflation}, 
Phys. Rev. D 101, 063535 (2020), arXiv:1907.02951 [astro-ph.CO].

\bibitem{AW}
A. Achucarro, Y. Welling, {\em Orbital Inflation: inflating along an angular isometry of field space}, arXiv:1907.02020 [hep-th].

\bibitem{MT}
M. Takada, {\em Can a galaxy redshift survey measure dark energy clustering?}, Phys. Rev. D74 (2006) 043505, arXiv:astro-ph/0606533.

\bibitem{HJ}
W. Hu, B. Jain, {\em Joint Galaxy-Lensing Observables and the Dark Energy}, Phys. Rev. D70 (2004) 043009, arXiv:astro-ph/0312395.

\bibitem{finalPl}
M. Tristram, A.J. Banday, M. Douspis, X. Garrido, K.M. Gorski, S. Henrot-Versille, S. Ilic, R. Keskitalo, G. Lagache, C.R. Lawrence, B. Partridge, D. Scott, {\em Cosmological parameters derived from the final (PR4) Planck data release}, 	arXiv:2309.10034 [astro-ph.CO].

\bibitem{HW}
J.-P. Hu, F.-Y. Wang, {\em Hubble Tension: The Evidence of New Physics}, Universe 9 (2023) 94, arXiv:2302.05709 [astro-ph.CO].

\bibitem{SV}
S. Vagnozzi, {\em Seven hints that early-time new physics alone is not sufficient to solve the Hubble tension}, Universe 9 (2023) 393, arXiv:2308.16628 [astro-ph.CO].

\bibitem{BCHCSJY}
A. Banerjee, H. Cai, L. Heisenberg, E. Colg\'ain, M. M. Sheikh-Jabbari, T. Yang, {\em Hubble Sinks In The Low-Redshift Swampland}, Phys. Rev. D 103 (2021) 081305, arXiv:2006.00244 [astro-ph.CO].

\bibitem{PZJ}
L. Pogosian, G.-B. Zhao, K. Jedamzik, {\em Recombination-independent determination of the sound horizon and the Hubble constant from BAO}, Astrophys. J. Lett. 904 (2020) L17, arXiv:2009.08455 [astro-ph.CO].

\bibitem{eBOSS}
eBOSS collaboration, {\em Completed SDSS-IV extended Baryon Oscillation Spectroscopic Survey: Cosmological implications from two decades of spectroscopic surveys at the Apache Point Observatory}, Phys. Rev. D 103 (2021) 083533, arXiv:2007.08991 [astro-ph.CO].

\bibitem{DescendingT}
K.C. Wong et al., {\em H0LiCOW-XIII. A 2.4\% measurement of $H_0$ from lensed quasars: 5.3$\sigma$ tension between early and late-Universe probes}, Mon. Not. R. Astron. Soc. 498 (2020) 1420-1439, arXiv:1907.04869 [astro-ph.CO].

\bibitem{DescendingT2}
C. Krishnan, E. Colg\'ain, Ruchika, A. Sen, M. Sheikh-Jabbari, T. Yang, {\em Is there an early Universe solution to Hubble tension?}, Phys. Rev. D 102 (2020) 103525, arXiv:2002.06044 [astro-ph.CO].

\bibitem{DescendingT3}
C. Krishnan, E. Colg\'ain, M. Sheikh-Jabbari, T. Yang, {\em Running Hubble tension and a H0 diagnostic}, Phys. Rev. D 103 (2021) 103509, arXiv:2011.02858 [astro-ph.CO].

\bibitem{DescendingT4}
M. Dainotti, B. De Simone, T. Schiavone, G. Montani, E. Rinaldi, G. Lambiase, {\em On the Hubble Constant Tension in the SNe Ia Pantheon Sample}, Astroph. J. 912 (2021) 150, arXiv:2103.02117 [astro-ph.CO].

\bibitem{DescendingT5}
N. Horstmann, Y. Pietschke, D. Schwarz, {\em Inference of the cosmic rest-frame from supernovae Ia}, Astron. Astrophys. 668 (2022) A34, arXiv:2111.03055 [astro-ph.CO].

\bibitem{DescendingT6}
M. Dainotti, B. De Simone, T. Schiavone, G. Montani, E. Rinaldi, G. Lambiase, M. Bogdan, S. Ugale, {\em On the Evolution of the Hubble Constant with the SNe Ia Pantheon Sample and Baryon Acoustic Oscillations: A Feasibility Study for GRB-Cosmology in 2030}, Galaxies 10 (2022) 24, arXiv:2201.09848 [astro-ph.CO]. 

\bibitem{DescendingT7}
E. Colg\'ain, M. Sheikh-Jabbari, R. Solomon, G. Bargiacchi, S. Capozziello, M.  Dainotti, D. Stojkovic, {\em Revealing Intrinsic Flat $\Lambda$CDM Biases with Standardizable Candles}, Phys. Rev. D 106 (2022) L041301, arXiv:2203.10558 [astro-ph.CO].

\bibitem{DescendingT8}
E. Colg\'ain, M. Sheikh-Jabbari, R. Solomon, M. Dainotti, D. Stojkovic, {\em Putting Flat $\Lambda$CDM In The (Redshift) Bin}, arXiv:2206.11447 [astro-ph.CO].

\bibitem{DescendingT9}
X. Jia, J. Hu, F. Wang, {\em Evidence of a decreasing trend for the Hubble constant}, Astron. Astrophys. 674 (2023) A45, arXiv:2212.00238 [astro-ph.CO].

\bibitem{MP}
V. Marra, L. Perivolaropoulos, {\em Rapid transition of $G_{eff}$ at $z_t \!\simeq \!0.01$ as a possible solution of the Hubble and growth tensions}, Phys.
Rev. D 104 (2021) L021303, arXiv:2102.06012 [astro-ph.CO].

\bibitem{HW2}
J. Hu, F. Wang, {\em Revealing the late-time transition of $H_0$: Relieve the Hubble crisis}, Mon. Not. R. Astron. Soc. 517 (2022) 576-581, arXiv:2203.13037 [astro-ph.CO].

\bibitem{SPYY}
M. Sharma, S. Pacif, G. Yergaliyeva, K. Yesmakhanova, {\em The Oscillatory Universe, phantom crossing and the Hubble tension}, Annals Phys. 454 (2023) 169345, arXiv:2205.13514 [gr-qc].

\bibitem{GSS}
M. Gangopadhyay, M. Sami, M. Sharma, {\em Phantom dark energy as a natural selection of evolutionary processes a la genetic algorithm and cosmological tensions}, Phys. Rev. D 108 (2023) 103526, arXiv:2303.07301 [astro-ph.CO].

\bibitem{CdR}
S. Capozziello and R. de Ritis, {\em Relation between the potential and nonminimal coupling in inflationary cosmology}, Phys. Lett. A177 (1993), 1.

\bibitem{CMRS}
S. Capozziello, G. Marmo, C. Rubano and P. Scudellaro, {\em Noether symmetries in Bianchi universes}, Int. J. Mod. Phys. D6 (1997) 491, gr-qc/9606050.

\bibitem{CNP}
S. Capozziello, S. Nesseris and L. Perivolaropoulos, {\em Reconstruction of the Scalar-Tensor Lagrangian from a LCDM Background and Noether Symmetry}, JCAP 0712 (2007) 009, arXiv:0705.3586 [astro-ph].

\bibitem{CDeF}
S. Capozziello and A. De Felice, {\em $f(R)$ Cosmology by Noether's symmetry}, JCAP 0808 (2008) 016, arXiv:0804.2163 [gr-qc].

\end{thebibliography}
\end{document}